\documentclass[structabstract]{aa}  
\usepackage{graphicx} 
\usepackage{txfonts}
\usepackage{natbib}
\usepackage{hyperref}
\usepackage{siunitx}  
\usepackage{xcolor}   

\newcommand{\GG}{\mbox{$G$}}
\newcommand{\GBP}{\mbox{$G_{\rm BP}$}}
\newcommand{\GRP}{\mbox{$G_{\rm RP}$}}

\newcommand{\nueff}{\mbox{$\nu_{\rm eff}$}}
\newcommand{\Zthree}{\mbox{$Z_{\rm EDR3}$}}

\newcommand{\sigmas}{\mbox{$\sigma_{\rm s}$}}
\newcommand{\sigmai}{\mbox{$\sigma_{\rm int}$}}
\newcommand{\sigmae}{\mbox{$\sigma_{\rm ext}$}}

\newcommand{\microas}{\mbox{$\mu$as}}
\newcommand{\microninv}{\mbox{$\mu$m$^{-1}$}}
\newcommand{\mci}[1]{\multicolumn{1}{c}{#1}}

\newcommand{\mciv}[1]{\multicolumn{4}{c}{#1}}

\newcommand{\pic}{\mbox{$\varpi_{\rm c}$}}

\newcommand{\pig}{\mbox{$\varpi_{\rm g}$}}

\newcommand{\VO}[1]{Villafranca~O-{#1}}

\begin{document}

   \title{An estimation of the \textit{Gaia} EDR3 parallax bias from stellar clusters and Magellanic Clouds data}
   \titlerunning{The \textit{Gaia} EDR3 parallax bias from stellar clusters and Magellanic Clouds data}

   \author{J. Ma\'{\i}z Apell\'aniz \inst{1}
           }
   \authorrunning{J. Ma\'{\i}z Apell\'aniz}

   \institute{Centro de Astrobiolog\'{\i}a, CSIC-INTA. Campus ESAC. 
              C. bajo del castillo s/n. 
              E-\num{28692} Villanueva de la Ca\~nada, Madrid, Spain\linebreak
              \email{jmaiz@cab.inta-csic.es} \\
             }

   \date{Received 4 October 2021 / Accepted 4 November 2021}

 
  \abstract
   {The early-third \textit{Gaia} data release (EDR3) parallaxes constitute the most detailed and accurate dataset that currently can be used to determine stellar distances in the solar 
    neighborhood.  Nevertheless, there is still room for improvement in their calibration and systematic effects can be further reduced in some circumstances.}
   {The aim of this paper is to determine an improved \textit{Gaia} EDR3 parallax bias as a function of magnitude, color, and ecliptic latitude using a single method 
    applied to stars in open clusters, globular clusters, the Large Magellanic Cloud (LMC), and the Small Magellanic Cloud (SMC).}
   {I study the behavior of the residuals or differences between the individual (stellar) parallaxes and the group parallaxes, which are assumed to be constant for the corresponding
    cluster or galaxy. This was done by first applying the Lindegren et al. (2021) zero point and then calculating a new zero point from the residuals of the first analysis.}
   {The Lindegren zero point shows very small residuals as a function of magnitude between individual and group parallaxes for $\GG> 13$ but significant ones for brighter stars, 
    especially blue ones. The new zero point reduces those residuals, especially in the $9.2 < \GG < 13$ range. The $k$ factor that is used to convert from catalog parallax
    uncertainties to external uncertainties is small (1.1-1.7) for $9.2 < \GG < 11$ and $\GG> 13$, intermediate (1.7-2.0) for $11 < \GG < 13$, and large ($>2.0$) for $\GG <9.2$.
    Therefore, significant corrections are needed to calculate distance uncertainties from \textit{Gaia}~EDR3 parallaxes for some stars. There is still room for improvement if
    future analyses add information from additional stellar clusters, especially for red stars with $\GG < 11$ and blue stars with $\GG <9.2$. I also calculated $k$ for stars 
    with RUWE values between 1.4 and 8.0 and for stars with six-parameter solutions, allowing for a correct estiimation of their uncertainties.}
   {}
   \maketitle
%

\section{Introduction}

$\,\!$\indent This is the second paper of a series on the validity of the parallaxes of the early third {\it Gaia} data release (EDR3; \citealt{Browetal21}), which was
presented on 3~December~2020 and included parallaxes for $\sim 1.5 \times 10^9$~sources. \citet{Lindetal21a}, from now on L21a, presents the astrometric solution for 
{\it Gaia} EDR3 and \citet{Lindetal21b}, from now on L21b, derived the parallax bias (or zero point, \Zthree) in the data as a function of magnitude (\GG, the primary very 
broadband optical photometry provided by {\it Gaia}), color (\nueff, the effective wavenumber, which for most well-behaved sources is a function of the \GBP$-$\GRP\ color 
provided by {\it Gaia}; see Fig.~2 in L21a), and ecliptic latitude ($\beta$). In the first paper of this series (\citealt{Maizetal21c}, from now on Paper~I) I used a 
variety of astrophysical sources to validate the results of L21a and L21b. In particular, I used {\it Gaia} EDR3 parallaxes for stars in globular clusters to test the 
\Zthree\ of L21b and determined that it works well for faint stars but that it can be improved for bright ones.

\begin{figure*}
 \centerline{\includegraphics[width=0.49\linewidth]{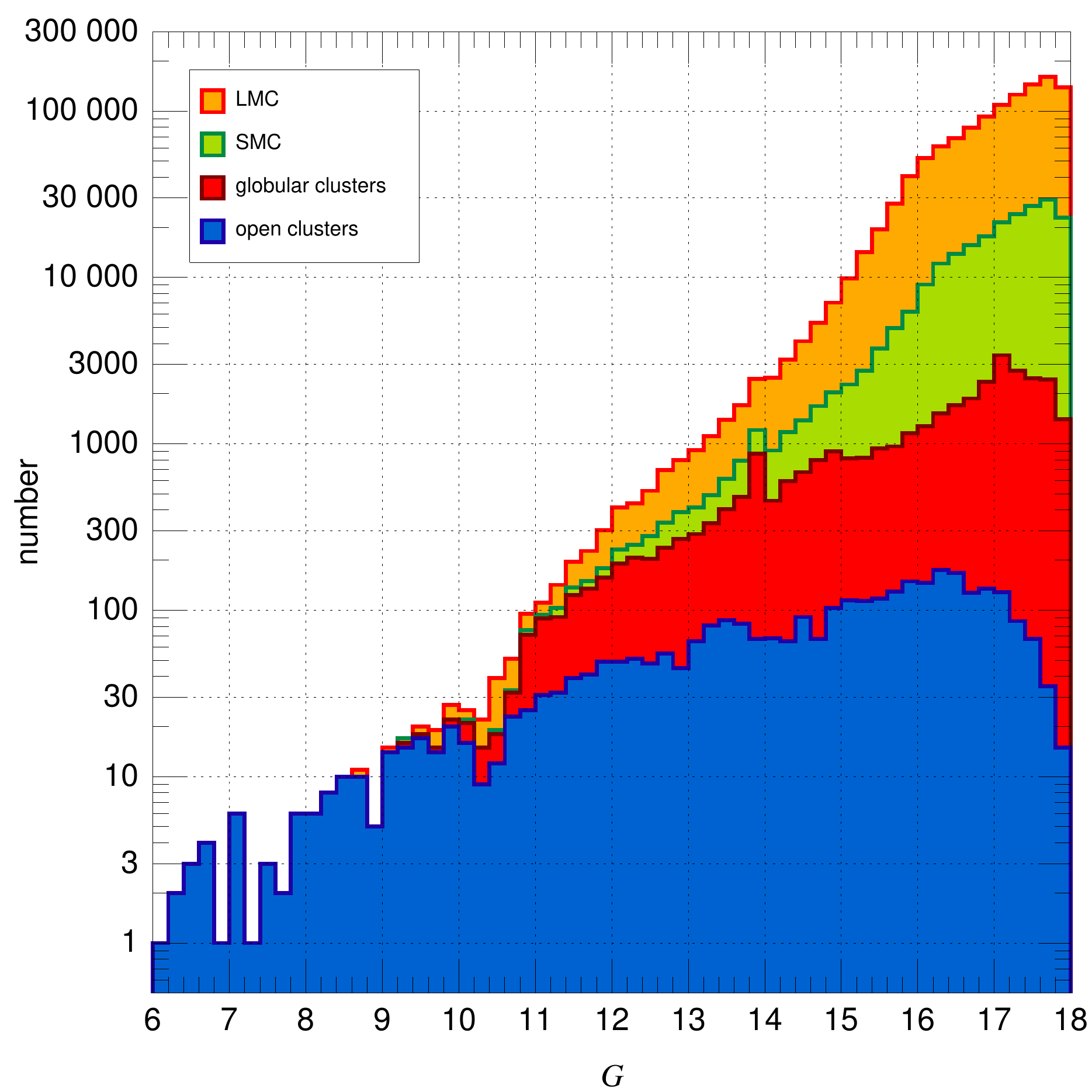} \
             \includegraphics[width=0.49\linewidth]{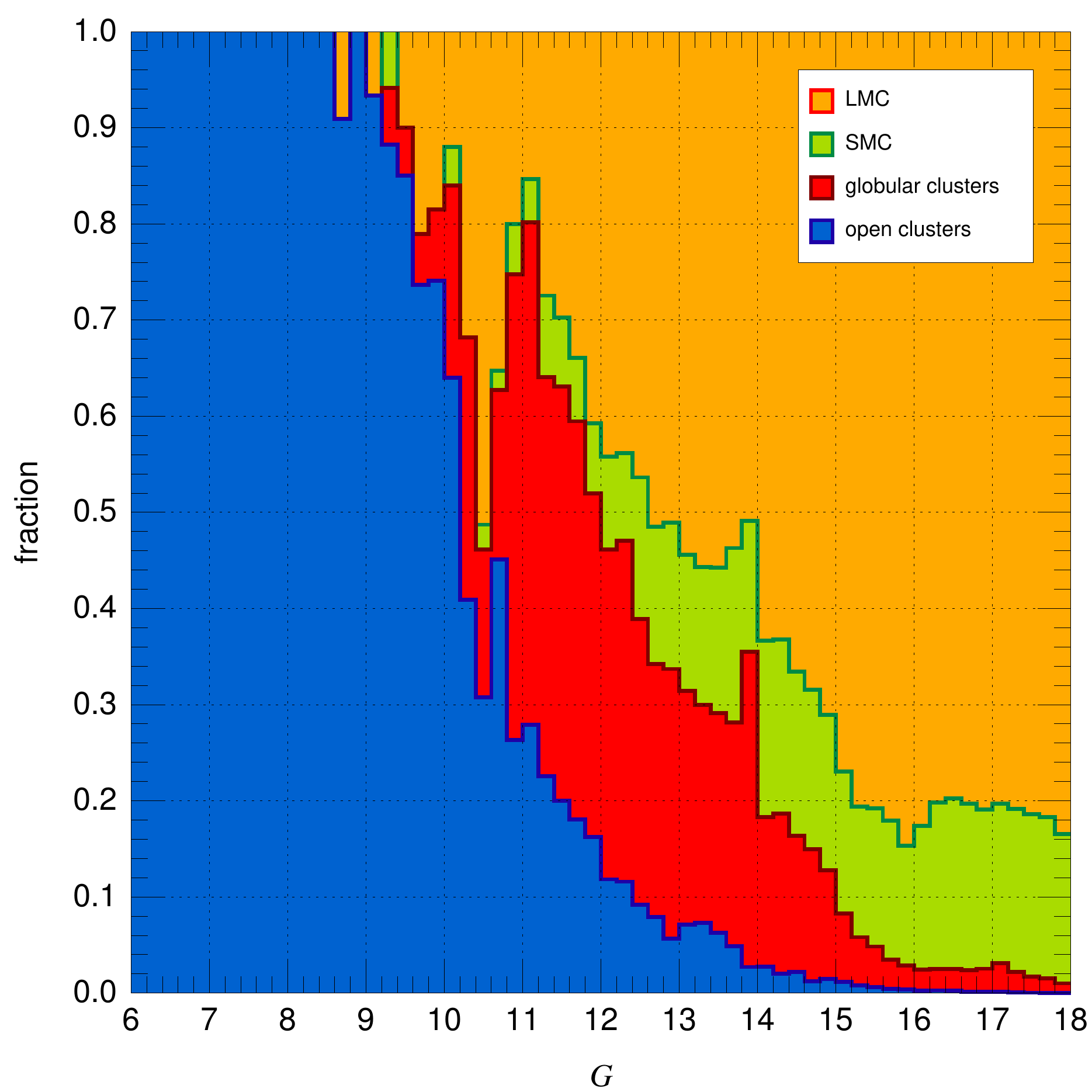}}
 \caption{(left) $G$ magnitude histogram for the samples in this paper using 0.2~mag bins and with a logarithmic vertical scale. I note that the distribution of each sample is 
          plotted above that of the following one, so the top line should be interpreted as the histogram for the total sample. (right) Fraction that each sample contributes 
          per 0.2~mag bin.}
 \label{ghisto}
\end{figure*}

The method used in Paper~I to validate \Zthree\ is based on the determination of distances to stellar groups from parallaxes established in \citet{Campetal19} and
\citet{Maiz19} and developed in \citet{Maizetal20b}, from now on Villafranca~I. That paper is the first one of a series on the Villafranca catalog of OB groups, Those OB groups, together
with those in the second paper of the series (Villafranca~II, described below), constitute an important part of the sample that is used in this paper. The analysis of \Zthree\ in
Paper~I uses the residual or difference between the individual parallaxes for each star and the parallax for the stellar group (in that paper, one of six globular clusters), \pig, to 
determine if the individual parallaxes require a correction (given by \Zthree) or, if the correction has already been calculated, whether it has the expected properties or not.
More specifically, I define the corrected individual parallaxes as:

\begin{equation}
\pic = \varpi - \Zthree
\label{pic}
\end{equation}

\noindent and the residual or difference with the group parallax as:

\begin{equation}
\Delta\varpi = \pic - \pig.
\label{Deltapi}
\end{equation}

\noindent The distribution of $\Delta\varpi$ normalized by its total or external uncertainty \sigmae\ should have a mean of zero and a standard deviation of one. The
external uncertainty was calculated as follows:

\begin{equation}
\sigmae = \sqrt{k^2 \sigma_{\rm int}^2 + \sigma_{\rm s}^2}, 
\label{sigmae}  
\end{equation}

\noindent where \sigmai\ is the internal (catalog) random uncertainty, \sigmas\ is the systematic uncertainty, and $k$ is a multiplicative constant that needs to be 
determined and that may depend on magnitude or other quantities. The value obtained for \sigmas\ in Paper~I was 10.3~\microas\ and that is the value that is used here. For
further details on the definitions, see Paper~I.

In this paper I present new estimations of the {\it Gaia}~EDR3 parallax bias and of $k$ based on the analysis of the parallaxes of stars that belong to open clusters (from
Villafranca~I~and~II), globular clusters (from Paper~I), and from the LMC and the SMC (from \citealt{Lurietal21}). The new \Zthree\ relies on the absolute QSO zero point of 
L21b but is otherwise independent of their calculation. The new estimations are applicable only to stars with five-parameter astrometric solutions in the case of \Zthree\ but for
$k$ I extend the analysis to six-parameter astrometric solutions. In the next section I describe the data and methods and in the following one I present and analyze the results. 
I conclude with a summary and possible ways to expand this work.

\begin{figure*}
 \centerline{\includegraphics[width=0.52\linewidth]{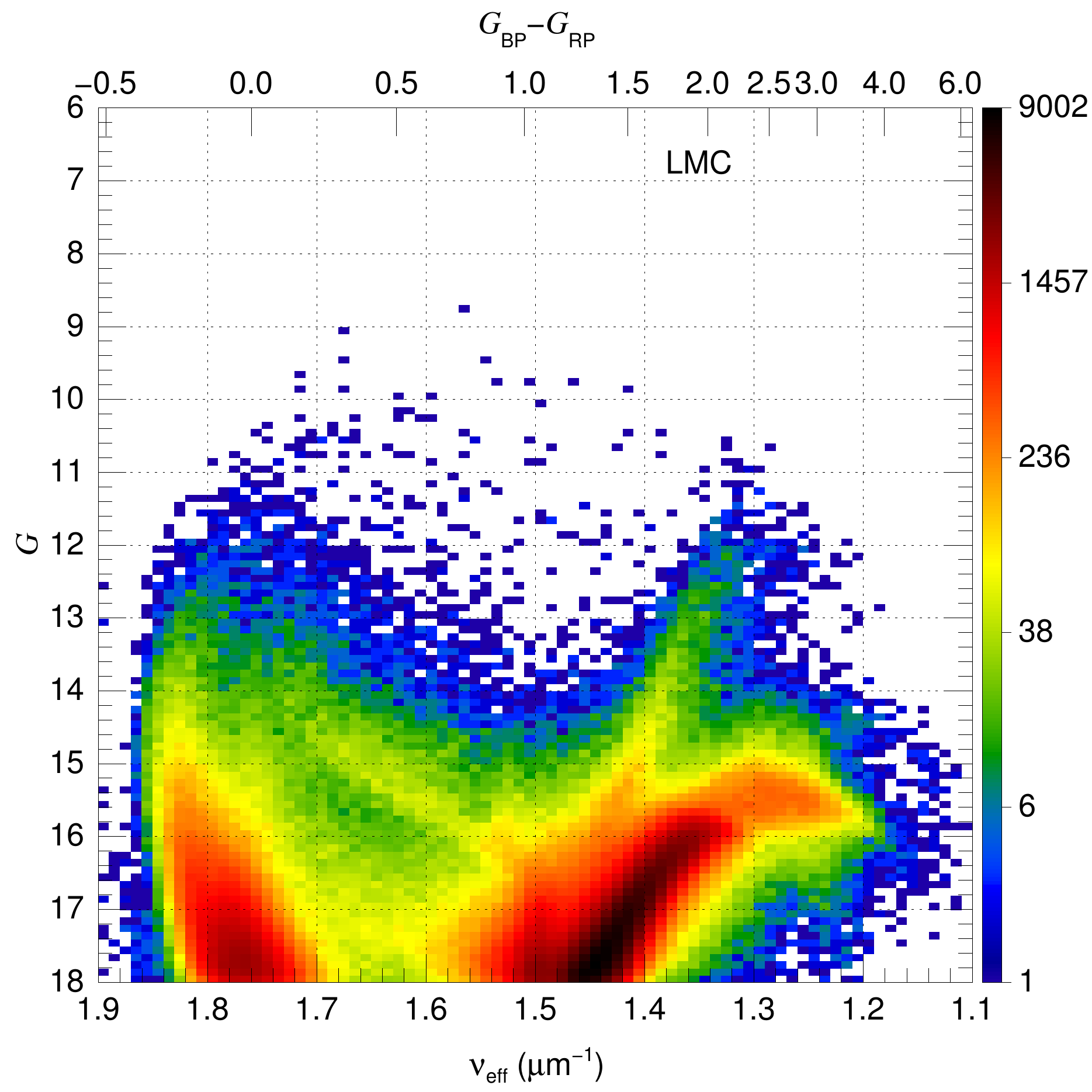}$\!\!$\includegraphics[width=0.52\linewidth]{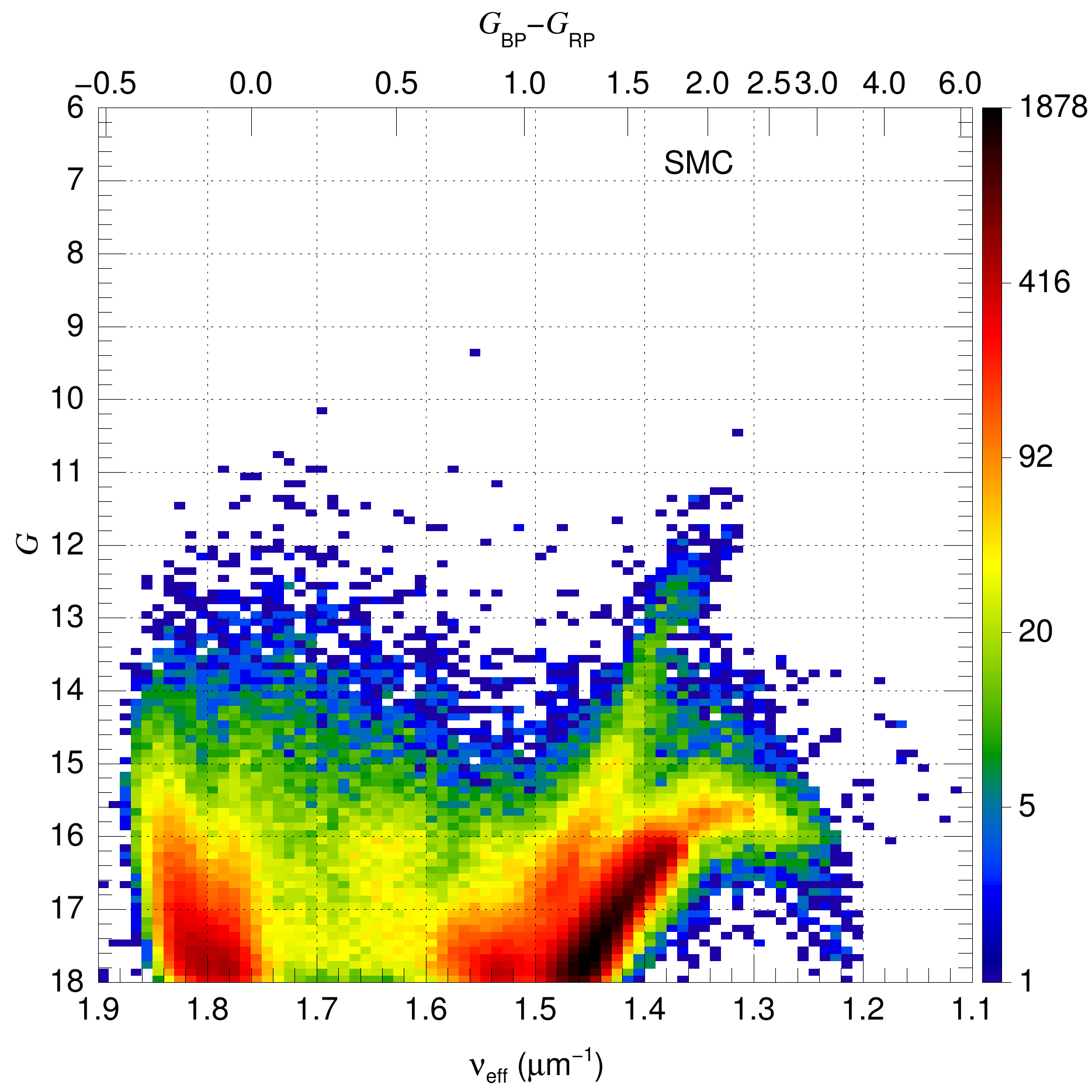}}
 \centerline{\includegraphics[width=0.52\linewidth]{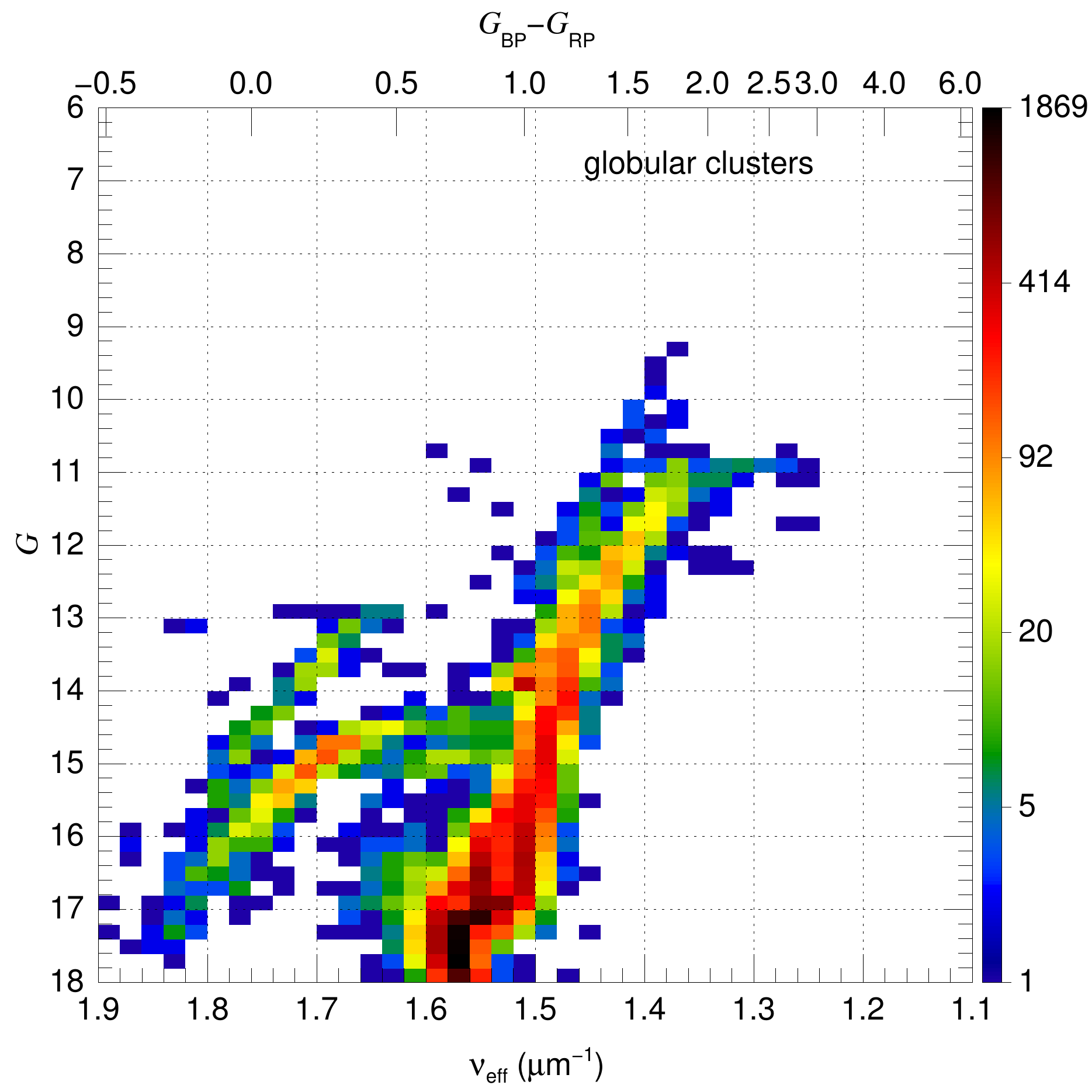}$\!\!$\includegraphics[width=0.52\linewidth]{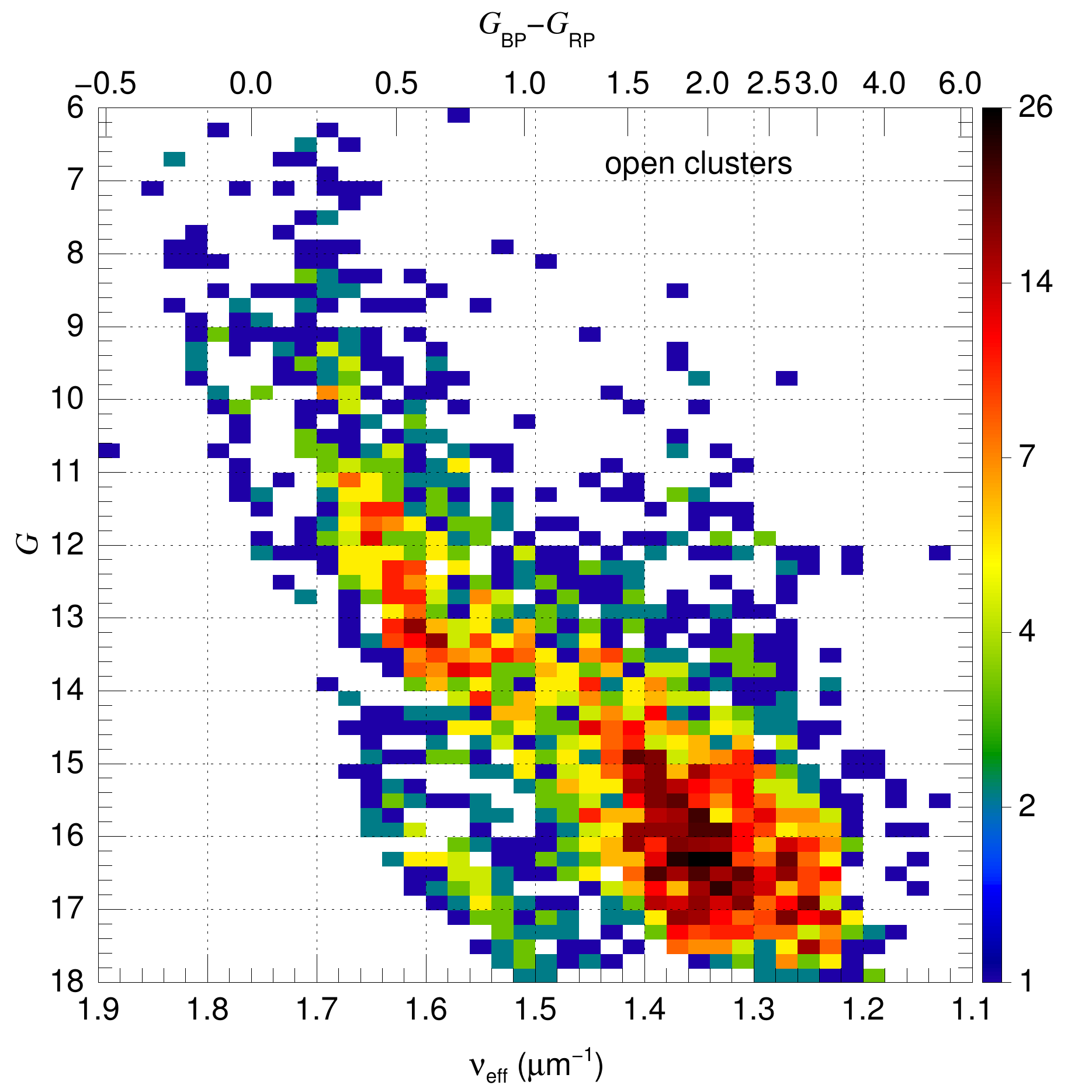}}
 \caption{CMDs for the LMC (upper left), SMC (upper right), globular cluster (lower left), and open cluster (lower right) samples. The intensity scale is
          logarithmic and the cell sizes are 0.01~\microninv~$\times$~0.1~mag (upper panels) and 0.02~\microninv~$\times$~0.2~mag (lower panels). The color bars at the right of each
          plot give the number of stars per cell, I note that they are very different from panel to panel. The upper x axes use Eqn.~4 of L21a to transform from \nueff\ to $\GBP-\GRP$.}
 \label{CMDs}
\end{figure*}

\section{Data and methods}

\subsection{How the sample was selected}          

$\,\!$\indent As mentioned above, the sample in this paper is a combination of stars with good-quality {\it Gaia}~EDR3 data from four types of objects: the LMC, the SMC, globular 
clusters, and open clusters. I first describe the selection process for the Magellanic Clouds and then I do it for the stellar clusters. 

For the LMC and the SMC I start with the sample obtained by \citet{Lurietal21}, who used an iterative procedure to eliminate nonmembers. I then restrict the sample to those objects
[a] within 10$^{\rm o}$ of the respective galaxy centers, [b] with RUWE~$<$~1.4 (Renormalized Unit Weight Error, see L21a), [c] with five-parameter {\it Gaia}~EDR3 solutions (see L21a), 
and [d] with $\sigmae < 100$~\microas\ using the $k$ and \sigmas\ values from paper I. This leaves us with a total of \num{989909} LMC stars and \num{196413} SMC stars with 
good-quality {\it Gaia}~EDR3 parallaxes. As there are few stars fainter than \GG~=~18~mag among those and because of the way the rest of the sample is selected, I add the additional 
condition [e] with $\GG < 18$~mag, leaving us with \num{950696} LMC stars and \num{192266} SMC stars.

For the globular clusters I use the sample from Paper I that consists of stars from six such systems ($\omega$~Cen, 47~Tuc, NGC~6752, M5, NGC~6397, and M13). The
selection of the stars was performed applying the same [b] to [e] conditions as in the previous paragraph, with the total number of stars for the six globular clusters 
being \num{30577}, ranging from the \num{1154} in M5 to the \num{14606} in 47~Tuc (I note that Table~5 in Paper~I includes stars with six-parameter solutions, which are excluded here). 
The sample in each cluster is selected by position and proper motion and applying a 4$\sigma$ cut in normalized parallax (see Paper I for details). The selection technique can be 
described as a simplified version as the one used by \citet{Lurietal21}, as globular clusters are more simple systems than the MCs and are also located at closer distances, which 
makes it easier to eliminate contaminants.

The sample for the open clusters\footnote{I use the term ``open cluster'' to refer to the OB groups defined in Villafranca~I~and~II even though some of them are not strictly bound
systems. However, for the purposes of this paper that is irrelevant, as their spread in distances introduces a dispersion in their parallaxes that is small in comparison with the
effects I am trying to measure.} consists of stars from 26 such systems, named as \VO{001} to \VO{026}. The first sixteen of those were originally defined in Villafranca~I
and the next ten were added in 
{Villafranca~II \citep{Maizetal21e}.} 
In that second paper the main criterion for adding OB groups to the list was precisely 
their usefulness for the analysis here, namely, the addition of a large number of stars per cluster with a high degree of certainty in membership. The sixteen Villafranca~I OB groups 
were originally analyzed with {\it Gaia}~DR2 astrometry but they were re-analyzed with {\it Gaia}~EDR3 astrometry (together with the ten new groups) in Villafranca~II.
For the North America nebula I use \VO{014}~NW as the cluster. The selection process of the 
{sample in each open cluster is the same as that in Villafranca~I~and~II and} 
is similar to that of the globular clusters with
two general differences: an additional cut using a displaced isochrone in the CMD is applied and the cut in normalized parallax is applied at 3$\sigma$ instead of at 4$\sigma$. The 
reason behind the first difference is that OB groups are usually located close to the Galactic plane, where the field population is a stronger contaminant than for globular clusters. 
The second difference arises from the lower number of stars selected per cluster, which ranges from 11 (\VO{013}) to 482 (\VO{022}), for a total of 3155 stars in open clusters (I note 
that the numbers in Villafranca~I are from {\it Gaia}~DR2, not {\it Gaia}~EDR3, and that the statistics in both Villafranca papers also include stars with six-parameter solutions).
Additionally, eleven stars that satisfied all requirements except that of the normalized parallax were added by hand because they are bright stars that were excluded by a small margin 
and additional information suggests they are indeed cluster members. As we see below, their original exclusion was caused by the underestimation of $k$ for bright stars in Paper I. 
The eleven stars are listed in Table~\ref{addedbyhand}.

\begin{table}
\caption{Stars in open clusters manually added to the sample.}
\centerline{
\addtolength{\tabcolsep}{-1mm}
\begin{tabular}{cclrc}
\hline
{\it Gaia} EDR3           & Clus. & Star name             & \mci{\GG} & \nueff       \\
ID                        & ID    &                       & (mag)     & (\microninv) \\
\hline
\num{5350363807177527680} & O-002 & ALS~\num{15863}       & 11.133    & 1.608        \\
\num{3326715714242517248} & O-016 & HD~\num{47755}        &  8.443    & 1.797        \\
\num{4146612425449995648} & O-019 & ALS~\num{15369}       & 11.217    & 1.486        \\
\num{3131327653265431552} & O-020 & HD~\num{46056}~B      & 10.895    & 1.661        \\
\num{5966509267005564288} & O-022 & HDE~\num{326331}~A    &  7.488    & 1.690        \\
\num{5966509438804267264} & O-022 & HD~\num{152314}~Aa,Ab &  7.802    & 1.680        \\
\num{5966508957767911040} & O-022 & HDE~\num{326330}      &  9.546    & 1.685        \\
\num{3017364063330465152} & O-023 & $\theta^1$~Ori~D      &  6.583    & 1.715        \\
\num{3017360348171372672} & O-023 & HD~\num{36982}        &  8.342    & 1.696        \\
\num{3017265961968363904} & O-023 & HD~\num{36939}        &  8.968    & 1.758        \\
\num{5350310927535609728} & O-025 & CPD~$-$59~2635        &  9.118    & 1.641        \\
\hline
\end{tabular}
\addtolength{\tabcolsep}{+1mm}
}
\label{addedbyhand}
\end{table}

\begin{figure}
 \centerline{\includegraphics[width=\linewidth]{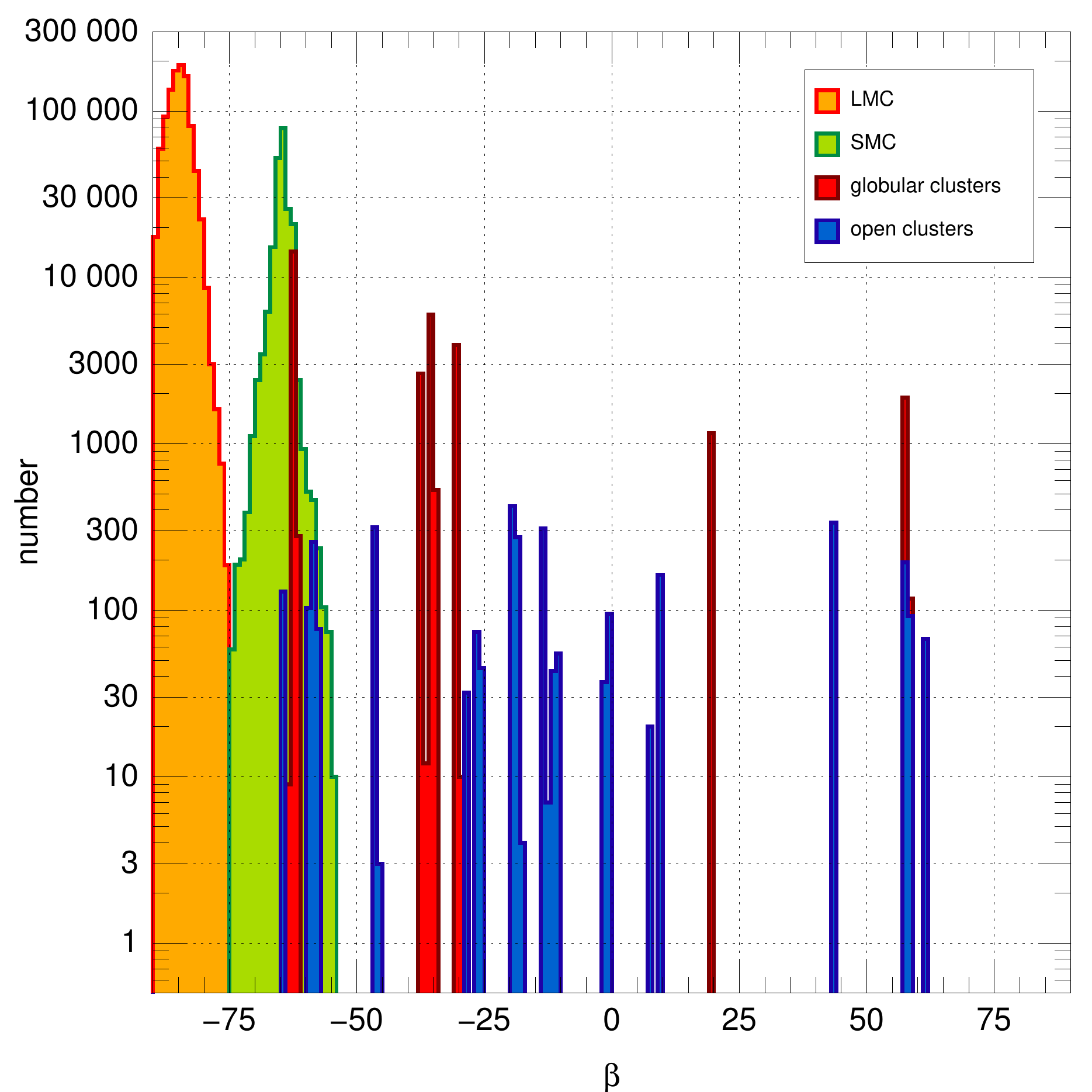}}
 \caption{$\beta$ 
          {(ecliptic latitude)} 
          histogram for the samples in this paper using 1$^{\rm o}$ bins and with a logarithmic vertical scale. I note that the distribution of each sample is 
          plotted above that of the following one, so the top line should be interpreted as the histogram for the total sample.}
 \label{lathisto}
\end{figure}

\subsection{Why the sample was selected}          

$\,\!$\indent The first goal in this paper is to derive a new parallax zero point for the five-parameter solutions in {\it Gaia}~EDR3 following the same parameter dependencies of L21b, that is:

\begin{equation}
\Zthree(G,\nu_{\rm eff},\beta) = \sum_{j=0}^{4} \sum_{k=0}^{2} q_{jk}(G)\, c_j(\nueff)\, b_k(\beta),
\label{Zdef}
\end{equation}

\noindent with the different terms explained in Appendix~A of L21b and section 2 of Paper I. In summary, there are eight possible terms for a given magnitude \GG: $q_{00}$, $q_{01}$, 
$q_{02}$ are the three color-independent $\beta$ terms; $q_{10}$, $q_{20}$, $q_{30}$, and $q_{40}$ are the three $\beta$-independent color terms (with the first one applying to the 
intermediate color range where most stars are located, the next two to red stars, and the last one to blue stars); and $q_{11}$ is the only term that depends on both color and 
$\beta$. Ideally, one should cover the \GG+\nueff+$\beta$ ranges of interest as thoroughly and uniformly as possible. In practice this is not possible for a number of reasons:

\begin{itemize}
 \item In general, faint stars are more common than bright ones.
 \item Single-age populations follow quasi-one-dimensional distributions (isochrones) in a CMD.
 \item Stars are not uniformly distributed over the whole sky.
 \item The presence of contaminants should be minimized.
\end{itemize}

The approach in this paper is to use the four different samples described above to try to cover the three-dimensional space of interest as best as possible. To describe how that is
done, I divide \GG\ in three ranges: faint ($13 < \GG < 18$), intermediate ($9.2 < \GG < 13$), and bright ($6 < \GG < 9.2$) and use Figs.~\ref{ghisto},~\ref{CMDs},~and~\ref{lathisto} 
to analyze how our sample is distributed.

The faint range is the better covered, as the LMC and SMC include objects of most magnitudes and colors and the small gaps left are well complemented with the other two samples. The
issue with the LMC and SMC is that they are both located near the south ecliptic pole, so clusters are needed to extend the solution to other latitudes (Fig.~\ref{lathisto}).

In the intermediate range the coverage is not as good and here is where the presence of different types of samples becomes even more useful. For the fainter end of the range the MCs 
are still the dominant contribution but only for blue and red colors, not intermediate ones. As we move to brighter magnitudes in this range, globular clusters become the most 
common population but only for red stars. Finally, for objects brighter than \GG~=~10~mag open clusters dominate but most of them are blue. 

The worst coverage of all is for bright stars. The sample is very small and it is composed almost exclusively of blue stars in open clusters.

In summary, I expect \Zthree\ to be better characterized for faint stars than for bright ones. However, I emphasize one important point of the technique employed in this paper.
Cluster (or galaxy) memberships are determined simultaneously for stars of very different magnitudes and colors and, given the behavior of the parallax uncertainties as a function of
magnitude, the largest weight in determining the distance (and, hence, establishing membership) is given to faint stars with $13 < \GG < 16$ for most systems. Therefore, stars in the
intermediate and bright ranges are anchored with respect to the better characterized faint stars. The reason for not doing a separate analysis for six-parameter solutions at this stage
is that the sample of such stars in the intermediate and faint ranges is small (Fig.~12 in L21b) but see below for the value of $k$.

\begin{figure*}
 \centerline{\includegraphics[width=0.50\linewidth]{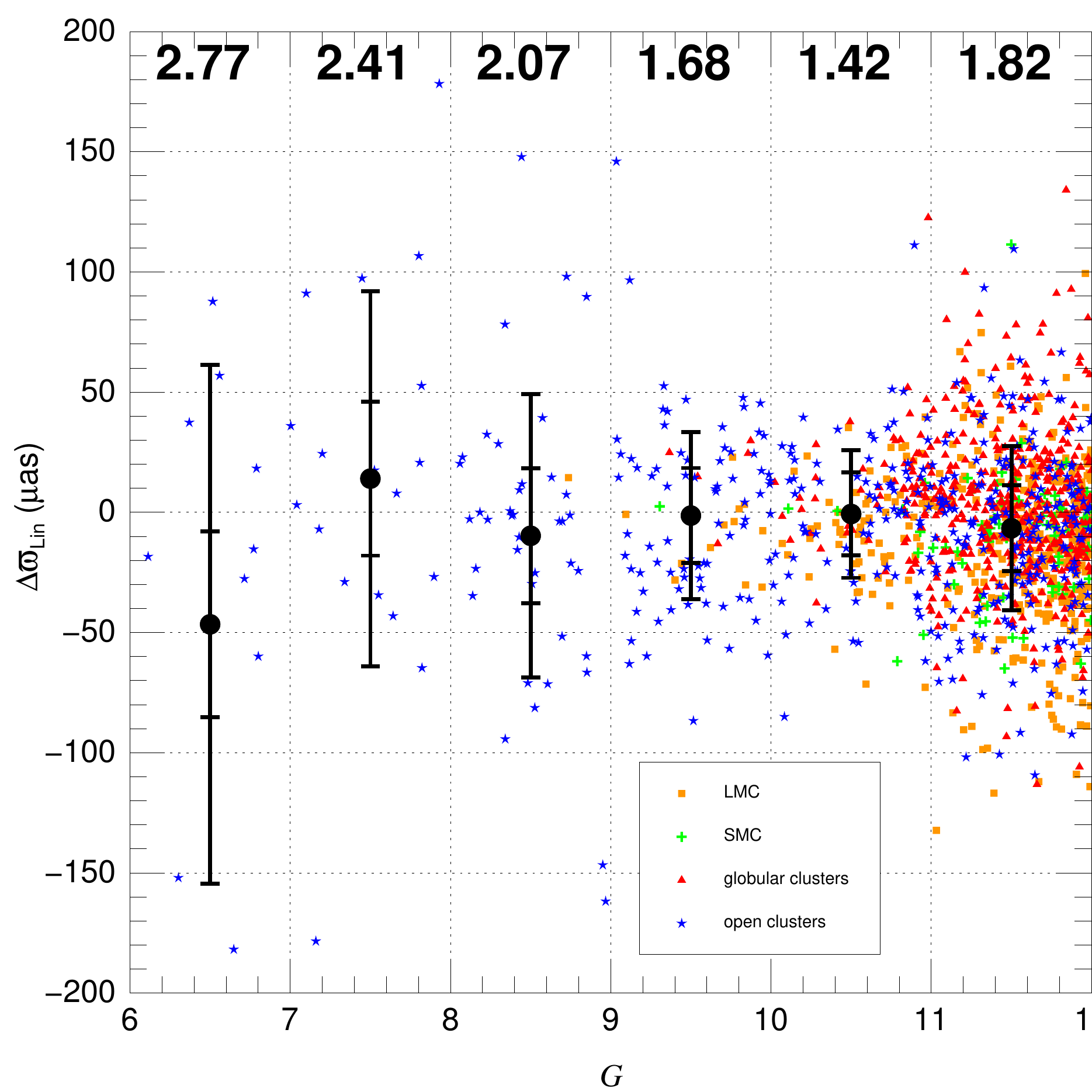}\includegraphics[width=0.50\linewidth]{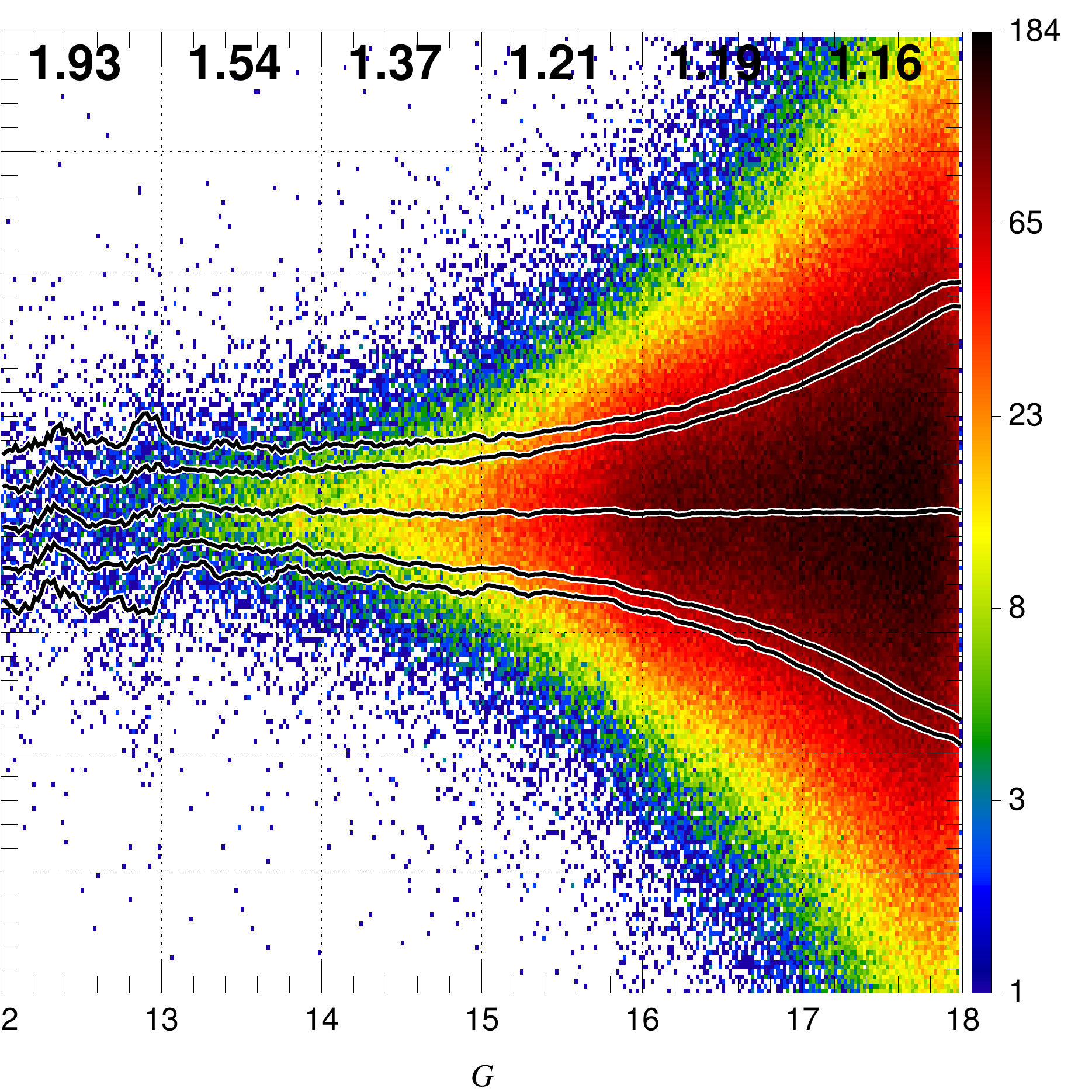}}
 \caption{Residuals using the Lindegren zero point, $\Delta\varpi_{\rm Lin}$, as a function of $G$. The plot is divided into two 
          panels due to the differences in the sample sizes for stars with $G$ between 6 and 12 mag (left panel) and stars with $G$ 
          {between 12 and 18 mag} 
          (right panel). In the left panel all stars are plotted individually with a color+symbol code used to differentiate between the four samples used in this 
          paper. In the right panel the stellar density is plotted combining all objects and using a logarithmic color scale, with the bar at the left indicating the 
          number of objects in each 0.02~mag~$\times$~2~$\mu$as~cell. The black points in the left panel show the average $\Delta\varpi_{\rm Lin}$ in each 
          magnitude bin. The error bars show the 
          {average of the parallax uncertainties} 
          (small values) and the dispersion of $\Delta\varpi_{\rm Lin}$ (large values), also in each 
          magnitude bin. In the right panel the points and error bars are substituted by lines displaying the same information. The text at the top of the panels 
          gives the value of $k$ in each magnitude bin as determined from the dispersion of $\Delta\varpi_{\rm Lin}$ and the average parallax uncertainty.}
 \label{gdpi_orig}
\end{figure*}

\begin{figure}
 \centerline{\includegraphics[width=\linewidth]{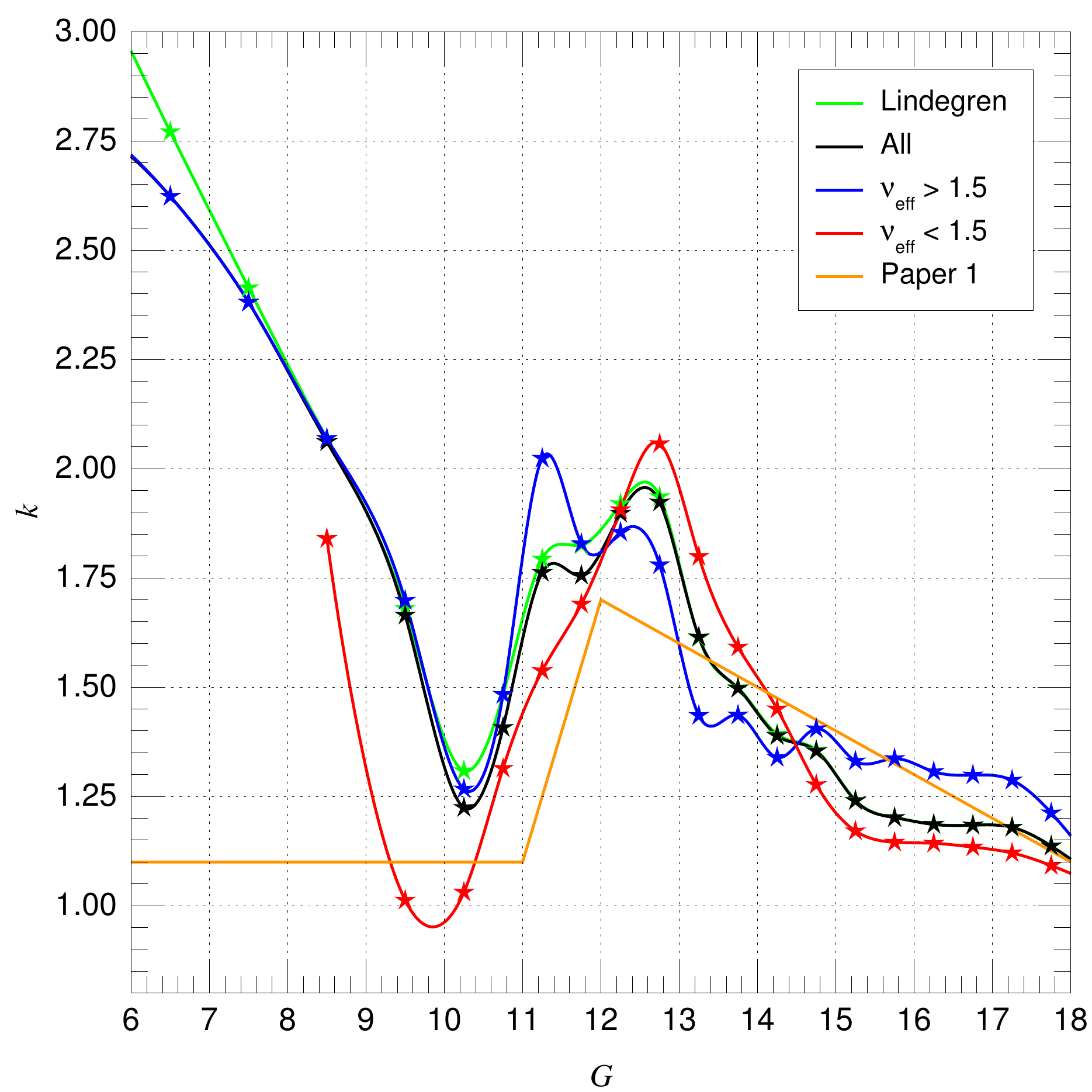}}
 \caption{$k$ as a function of magnitude for all colors using the Lindegren zero point (Table~\ref{stats_orig}) and for the three cases in Table~\ref{stats} using the results in
          this paper. The data points are calculated at 1~mag intervals for $\GG < 10$ and at 0.5~mag intervals for $\GG > 10$ and joined by a spline. The orange line shows the 
          approximation derived in Paper 1.}
 \label{k}
\end{figure}

\begin{figure*}
 \centerline{\includegraphics[width=0.49\linewidth]{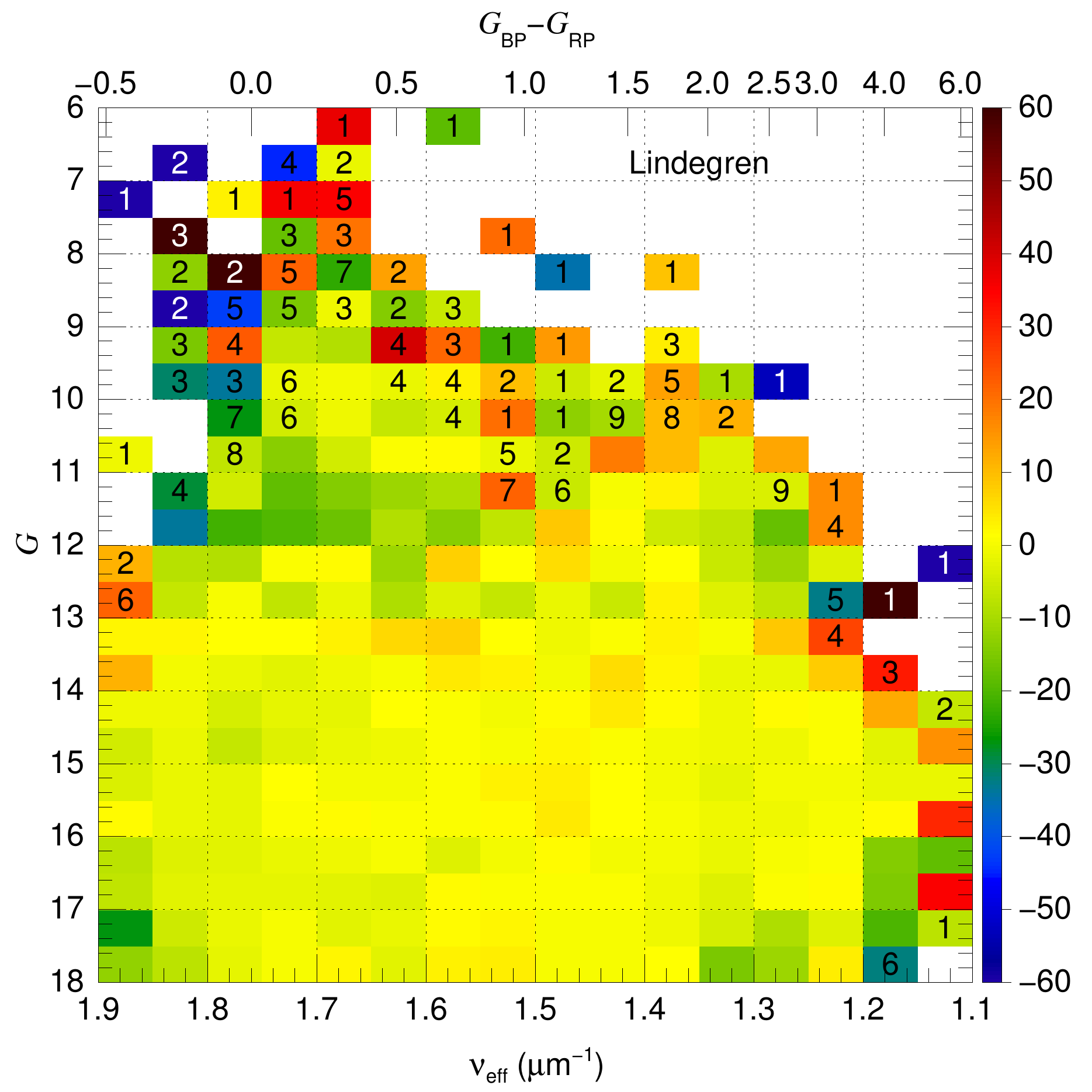}\
             \includegraphics[width=0.49\linewidth]{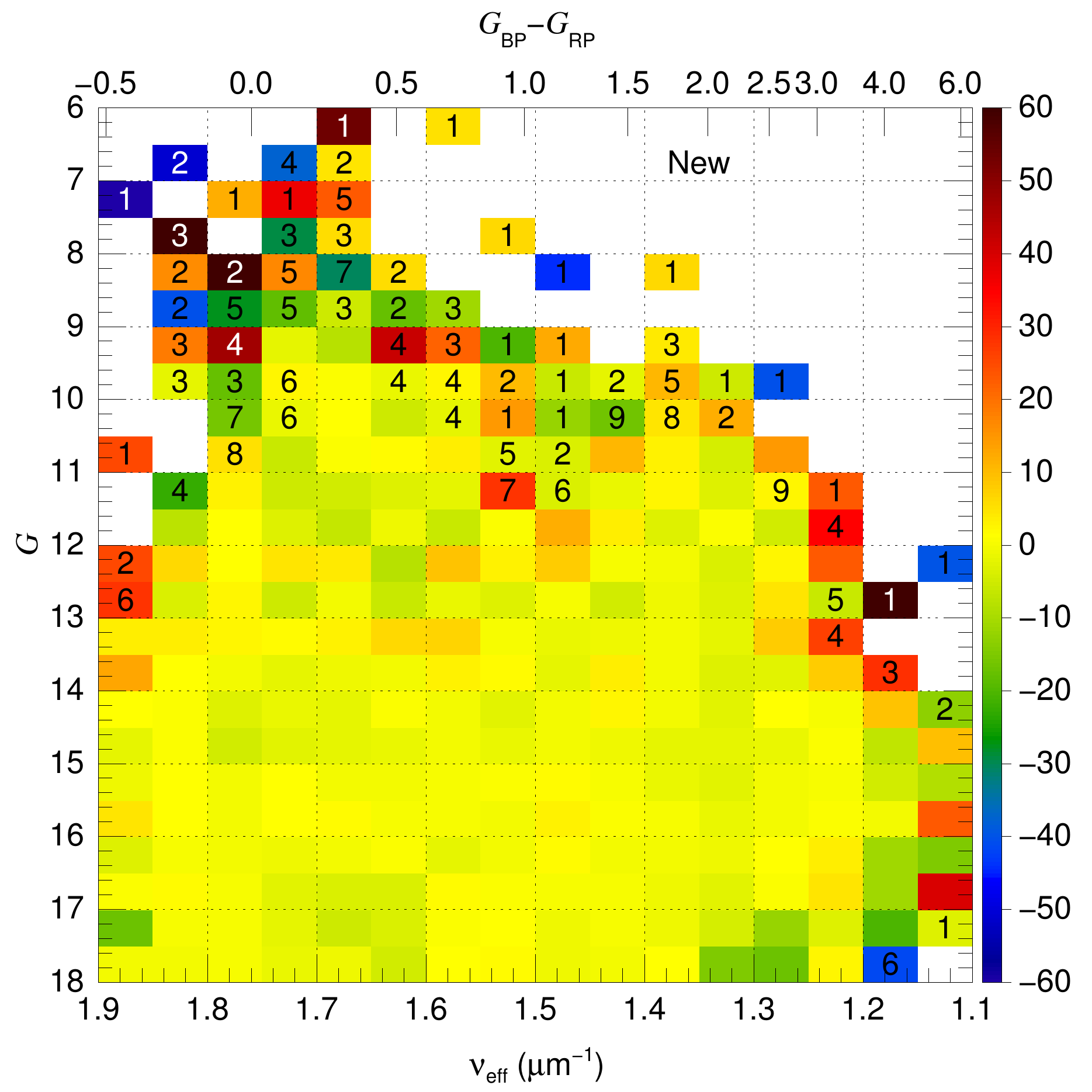}}
 \caption{Average residual as a function of \nueff\ and $G$ using the Lindegren zero point (left) and the one proposed here (right) for
          the full sample.  The left bar shows the scale in \microas. Each cell has a size of 0.05~\microninv~$\times$~0.5~mag. Cells with less than ten objects used to calculate 
          the average include the number of objects. The color scale is capped for values above 60~\microas\ or below $-$60~\microas\ for display purposes but some cells (seven
          in the left panel, and four in the right panel) are outside that range. The upper $x$ axes use Eqn.~4 of L21a to transform from \nueff\ to $\GBP-\GRP$.}
 \label{dpihisto}
\end{figure*}

\begin{table*}
\caption{Statistics (in \microas) by \GG\ magnitude and \nueff\ ranges using the Lindegren correction.
         For $\GG < 9$ the last columns are omitted as only two stars have $\nueff < 1.5$~\microninv.}
\centerline{
\begin{tabular}{crrrrrrrrrrrr}
\hline
          & \mciv{all} & \mciv{$\nueff > 1.5$} & \mciv{$\nueff < 1.5$} \\
\GG       & $N$    & $\overline{\Delta\varpi}_{\rm Lin}$ & $\sigma_{\Delta\varpi,{\rm Lin}}$ & \mci{$k_{\rm Lin}$} & $N$    & $\overline{\Delta\varpi}_{\rm Lin}$ & $\sigma_{\Delta\varpi,{\rm Lin}}$ & \mci{$k_{\rm Lin}$} & $N$    & $\overline{\Delta\varpi}_{\rm Lin}$ & $\sigma_{\Delta\varpi,{\rm Lin}}$ & \mci{$k_{\rm Lin}$} \\
\hline
  6.0- 7.0 & \num{11}     & $ -46.58$ & $ 107.83$ & 2.77 &          --- &       --- &       --- &  --- &          --- &       --- &       --- &  --- \\
  7.0- 8.0 & \num{18}     & $ +14.03$ & $  78.05$ & 2.41 &          --- &       --- &       --- &  --- &          --- &       --- &       --- &  --- \\
  8.0- 9.0 & \num{40}     & $  -9.75$ & $  58.97$ & 2.07 &          --- &       --- &       --- &  --- &          --- &       --- &       --- &  --- \\
  9.0-10.0 & \num{98}     & $  -1.35$ & $  34.80$ & 1.68 & \num{84}     & $  -1.82$ & $  36.64$ & 1.70 & \num{14}     & $  +1.45$ & $  21.35$ & 1.31 \\
 10.0-10.5 & \num{154}    & $  -3.77$ & $  24.30$ & 1.31 & \num{113}    & $  -5.25$ & $  26.06$ & 1.36 & \num{41}     & $  +0.31$ & $  18.28$ & 1.02 \\
 10.5-11.0 & \num{170}    & $  +0.82$ & $  27.27$ & 1.48 & \num{94}     & $  -4.42$ & $  28.07$ & 1.54 & \num{76}     & $  +7.30$ & $  24.93$ & 1.33 \\
 11.0-11.5 & \num{340}    & $  -3.48$ & $  36.11$ & 1.79 & \num{142}    & $ -10.24$ & $  41.09$ & 2.03 & \num{198}    & $  +1.36$ & $  31.29$ & 1.55 \\
 11.5-12.0 & \num{637}    & $  -8.29$ & $  32.93$ & 1.83 & \num{242}    & $ -16.98$ & $  35.42$ & 1.85 & \num{395}    & $  -2.96$ & $  30.13$ & 1.73 \\
 12.0-12.5 & \num{1098}   & $  -2.51$ & $  33.16$ & 1.92 & \num{394}    & $  -3.74$ & $  37.77$ & 1.88 & \num{704}    & $  -1.83$ & $  30.29$ & 1.93 \\
 12.5-13.0 & \num{1768}   & $  -2.89$ & $  36.84$ & 1.94 & \num{758}    & $  -4.35$ & $  38.58$ & 1.79 & \num{1010}   & $  -1.79$ & $  35.46$ & 2.07 \\
 13.0-13.5 & \num{2712}   & $  +1.95$ & $  26.74$ & 1.61 & \num{1380}   & $  +2.77$ & $  25.11$ & 1.44 & \num{1332}   & $  +1.10$ & $  28.31$ & 1.80 \\
 13.5-14.0 & \num{4877}   & $  +0.78$ & $  26.36$ & 1.50 & \num{2886}   & $  +0.42$ & $  26.08$ & 1.44 & \num{1991}   & $  +1.30$ & $  26.77$ & 1.59 \\
 14.0-14.5 & \num{7607}   & $  +0.26$ & $  28.22$ & 1.39 & \num{4060}   & $  -1.45$ & $  28.11$ & 1.34 & \num{3547}   & $  +2.22$ & $  28.23$ & 1.45 \\
 14.5-15.0 & \num{14602}  & $  -0.95$ & $  31.35$ & 1.36 & \num{7716}   & $  -2.05$ & $  33.79$ & 1.41 & \num{6886}   & $  +0.29$ & $  28.32$ & 1.28 \\
 15.0-15.5 & \num{32994}  & $  -0.10$ & $  32.73$ & 1.24 & \num{11952}  & $  -0.26$ & $  36.92$ & 1.33 & \num{21042}  & $  -0.01$ & $  30.08$ & 1.17 \\
 15.5-16.0 & \num{78681}  & $  +0.15$ & $  37.68$ & 1.20 & \num{19235}  & $  -0.20$ & $  43.98$ & 1.34 & \num{59446}  & $  +0.27$ & $  35.40$ & 1.15 \\
 16.0-16.5 & \num{146663} & $  -0.49$ & $  45.41$ & 1.19 & \num{32112}  & $  -2.00$ & $  52.78$ & 1.31 & \num{114551} & $  -0.07$ & $  43.11$ & 1.14 \\
 16.5-17.0 & \num{206874} & $  -0.20$ & $  57.49$ & 1.18 & \num{54569}  & $  -1.25$ & $  65.60$ & 1.30 & \num{152305} & $  +0.17$ & $  54.29$ & 1.13 \\
 17.0-17.5 & \num{303660} & $  -0.08$ & $  73.64$ & 1.18 & \num{97619}  & $  -0.75$ & $  81.98$ & 1.29 & \num{206041} & $  +0.24$ & $  69.33$ & 1.12 \\
 17.5-18.0 & \num{373782} & $  +0.20$ & $  89.80$ & 1.14 & \num{132270} & $  -0.14$ & $  96.33$ & 1.21 & \num{241512} & $  +0.39$ & $  86.01$ & 1.09 \\
\hline
\end{tabular}
}
\label{stats_orig}
\end{table*}

\begin{figure*}
 \centerline{\includegraphics[width=0.49\linewidth]{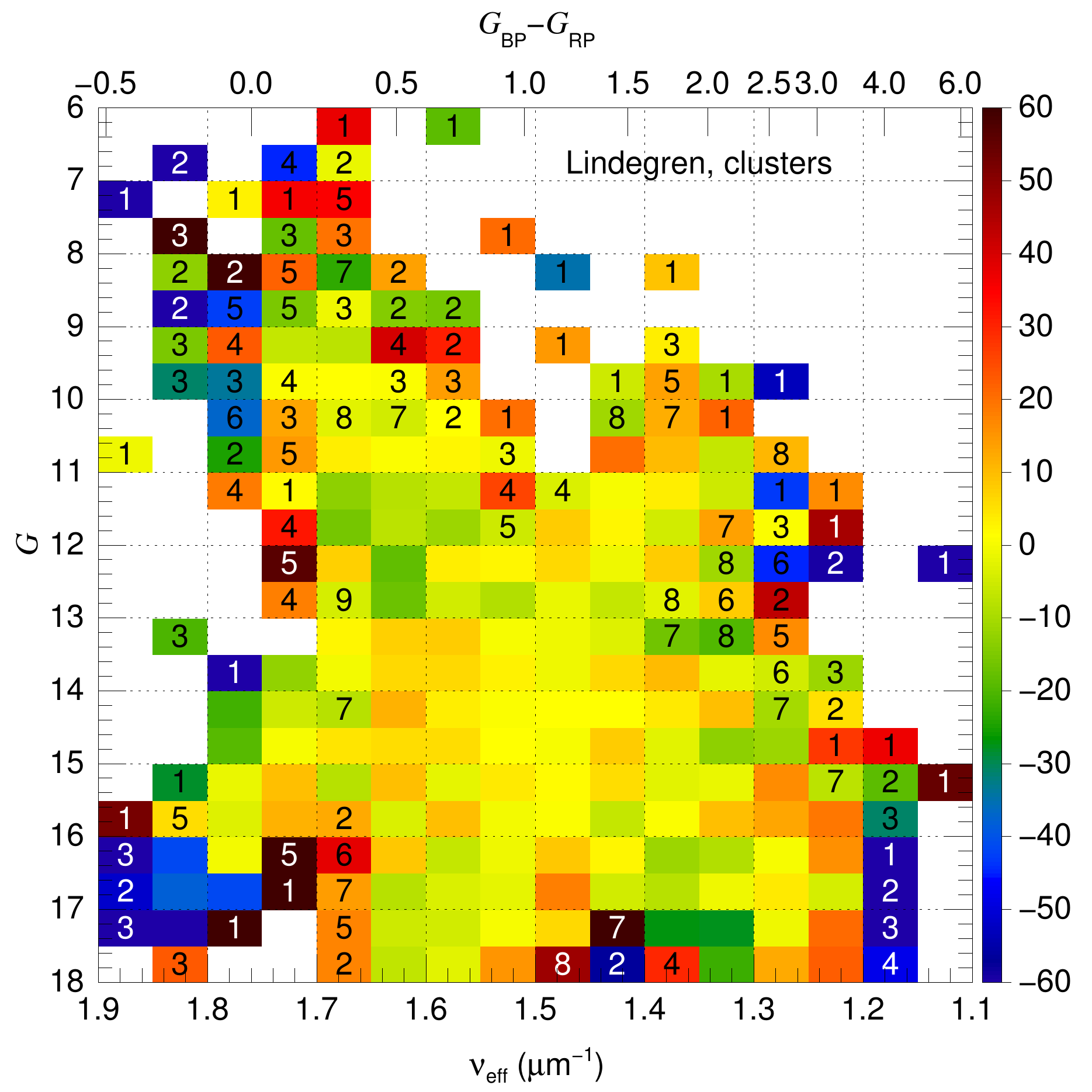}\
             \includegraphics[width=0.49\linewidth]{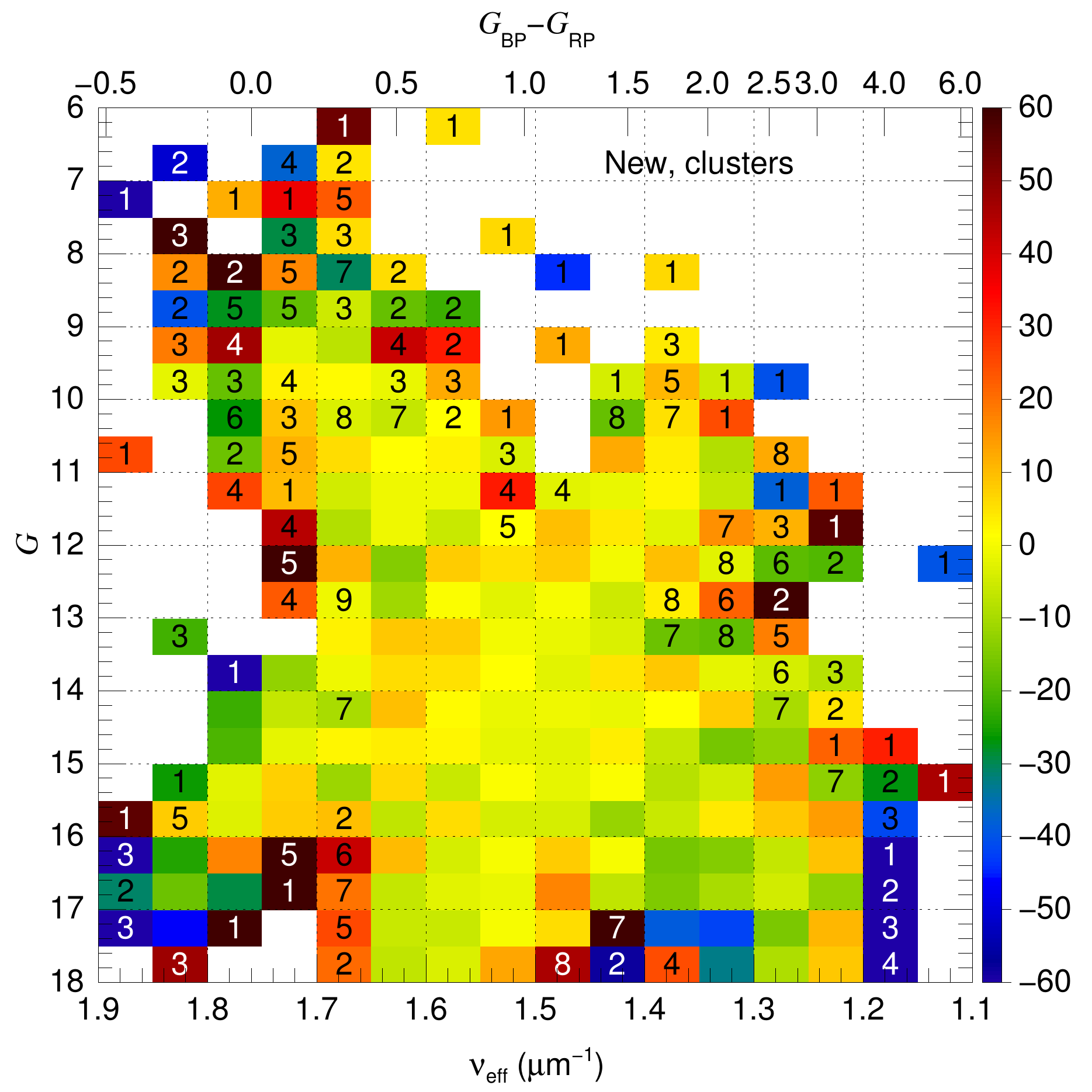}}
 \caption{Same as Fig.~\ref{dpihisto}, but using only the cluster sample. Here the number of cells outside the range is 17 in the left panel and 16 in the right panel.}
 \label{dpihistoclus}
\end{figure*}

\begin{figure*}
 \centerline{\includegraphics[width=0.49\linewidth]{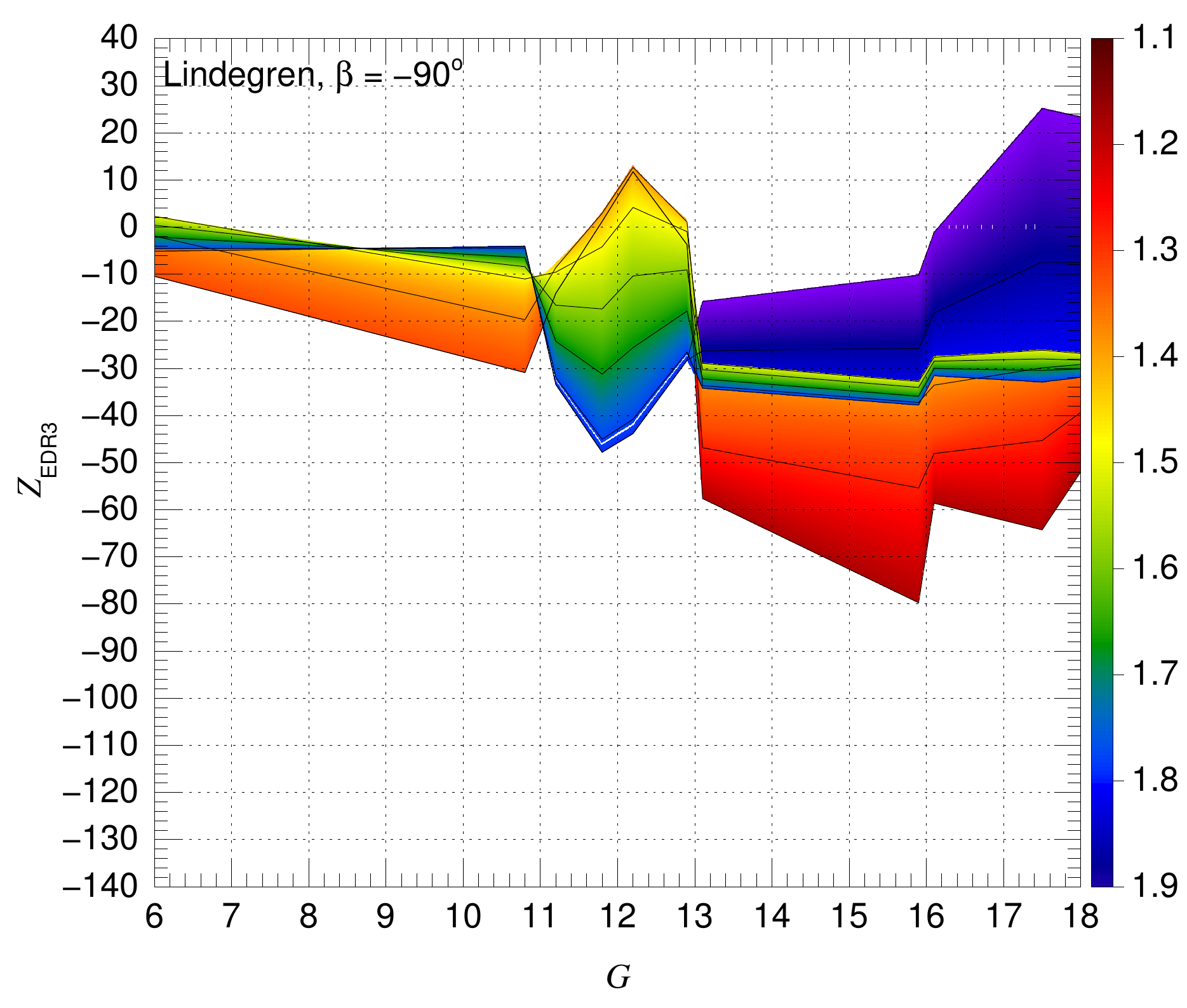}\
             \includegraphics[width=0.49\linewidth]{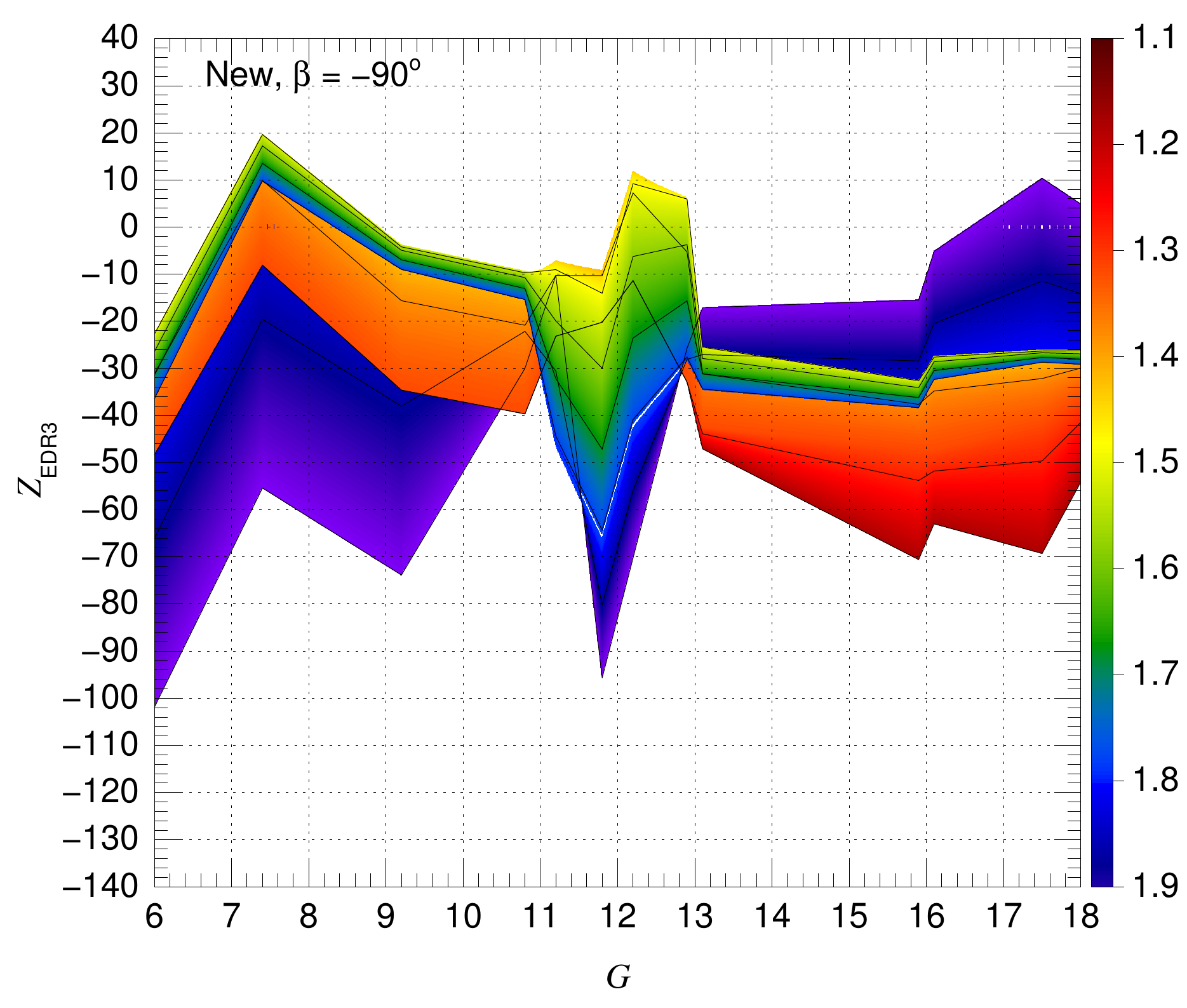}}
 \centerline{\includegraphics[width=0.49\linewidth]{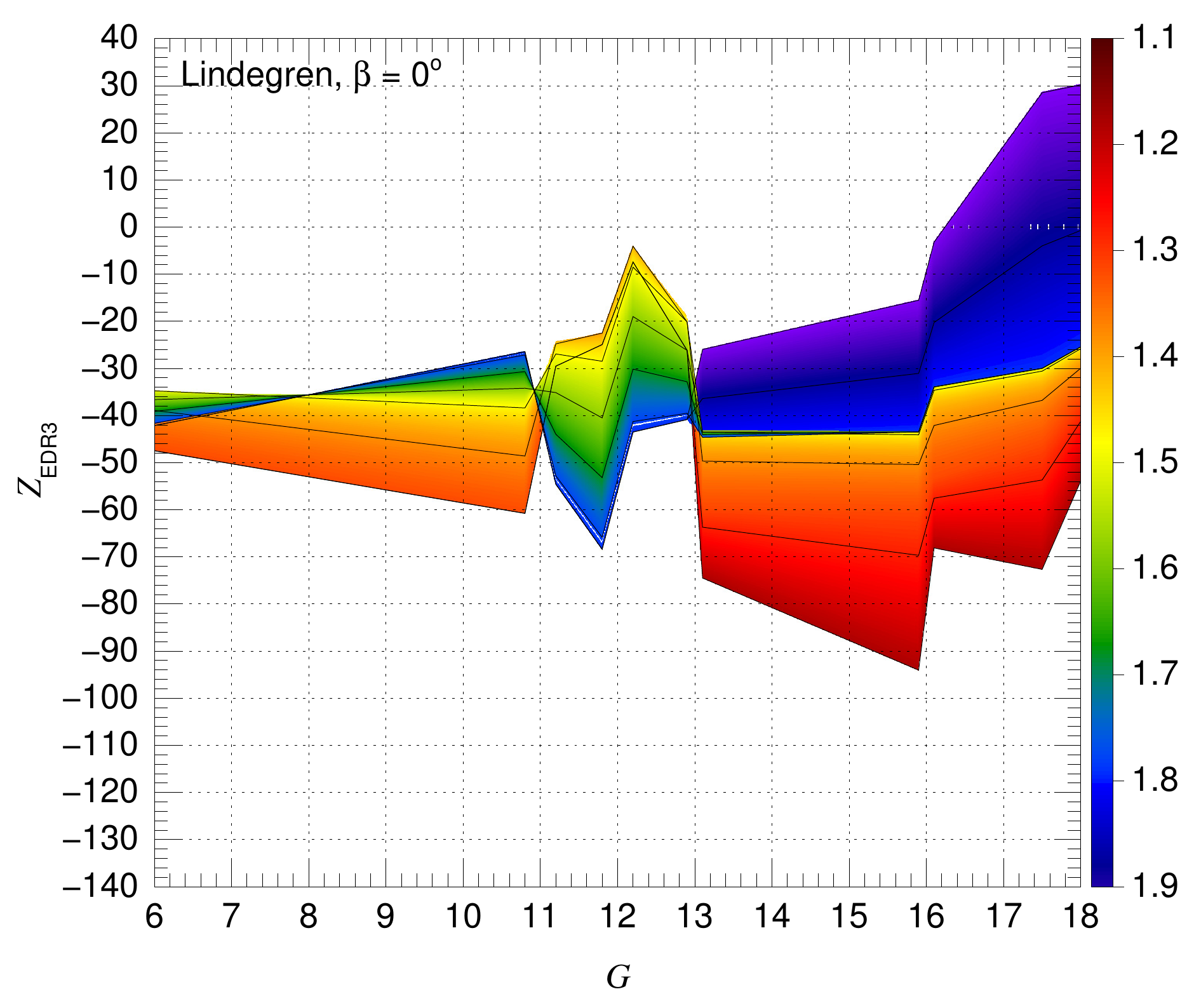}\
             \includegraphics[width=0.49\linewidth]{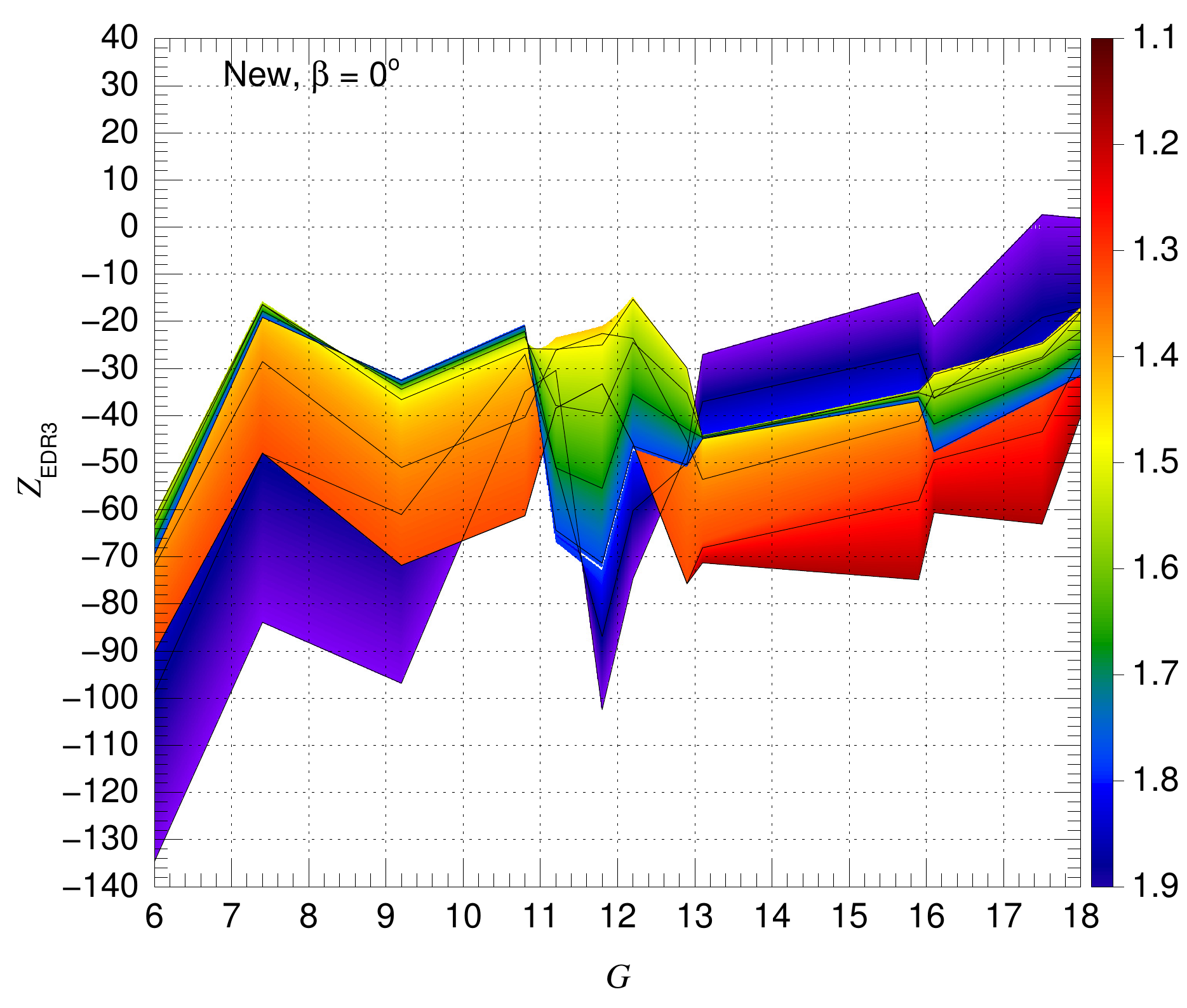}}
 \centerline{\includegraphics[width=0.49\linewidth]{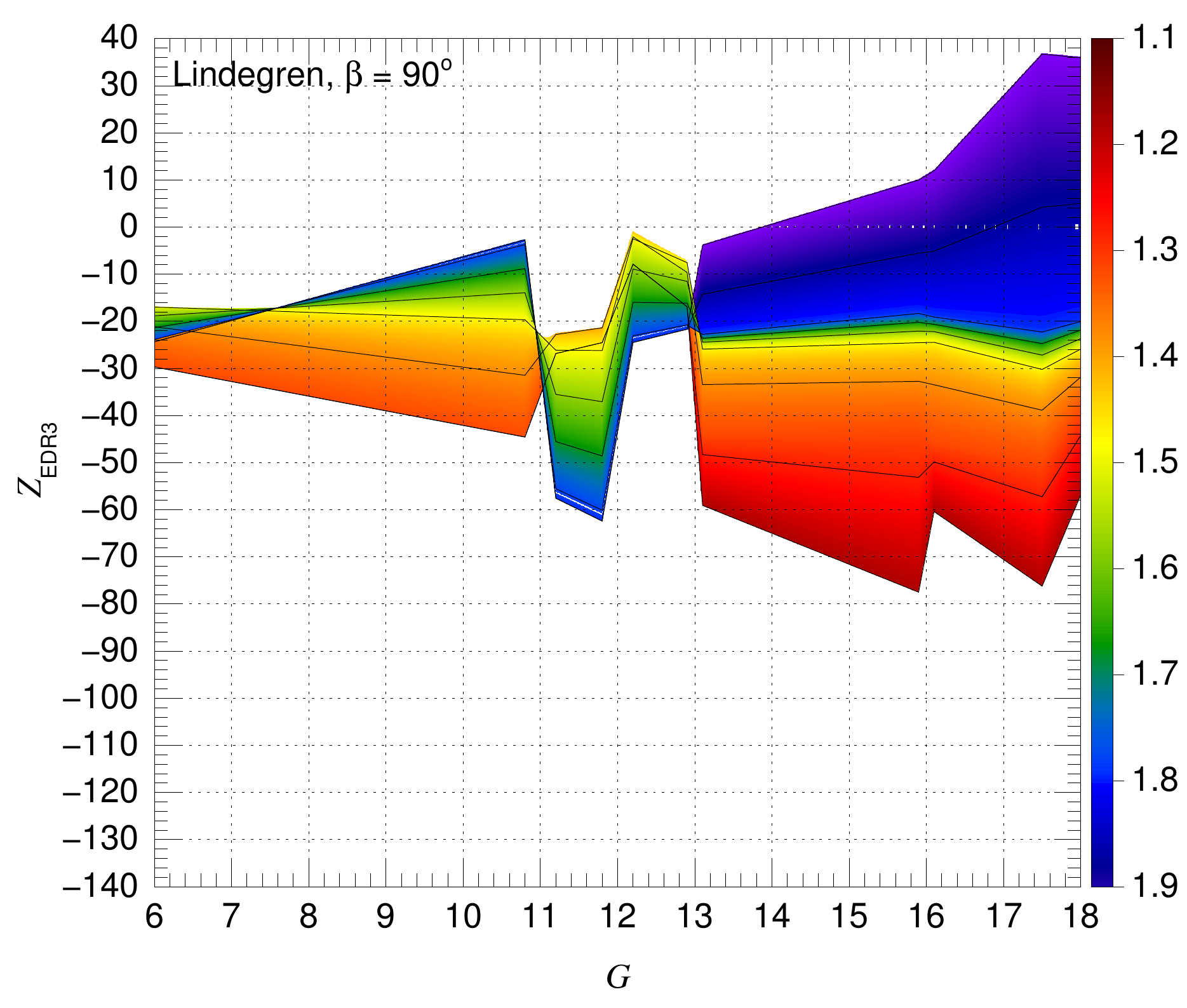}\
             \includegraphics[width=0.49\linewidth]{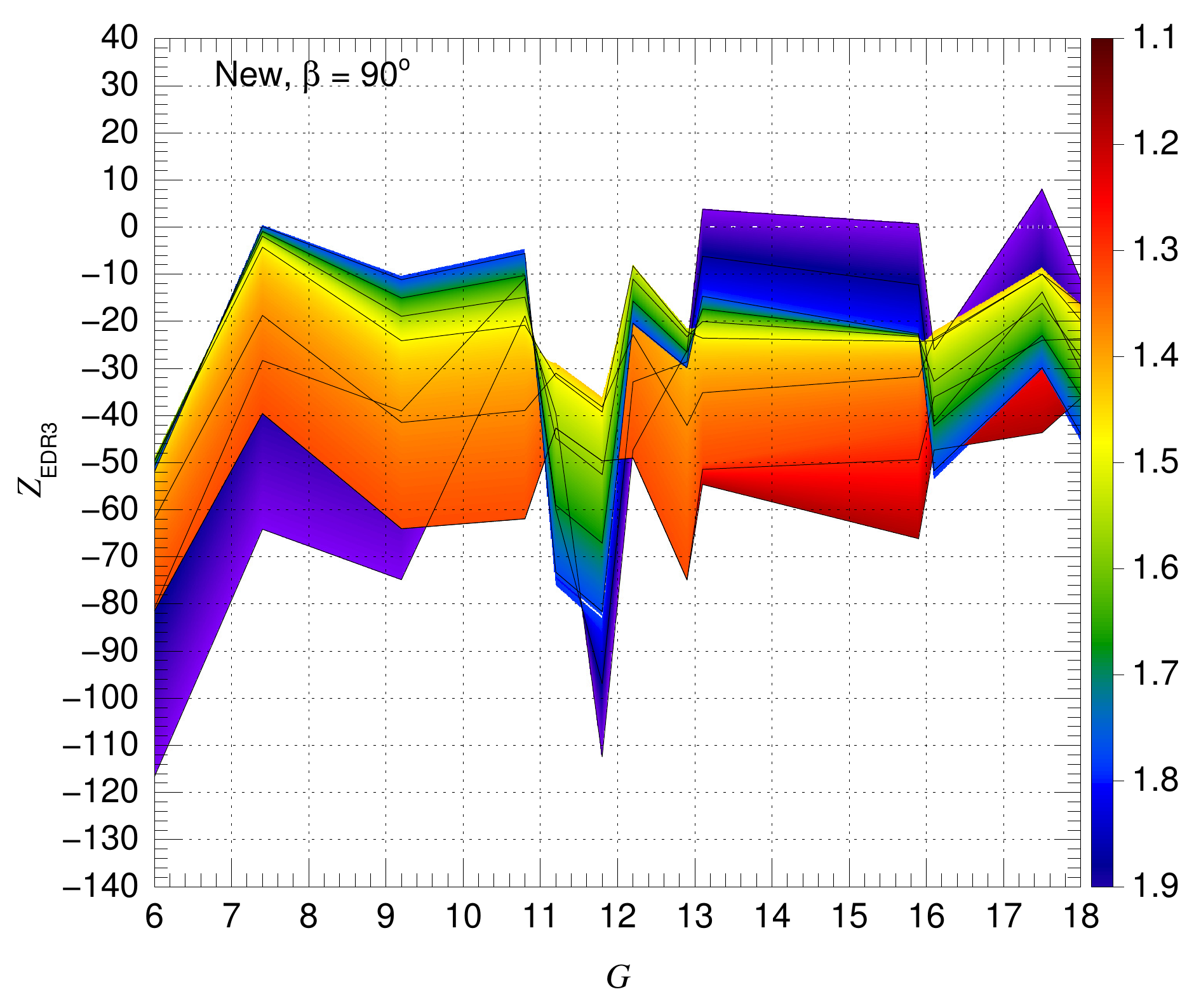}}
 \caption{\Zthree\ as a function of \GG\ (horizontal axes) and \nueff\ (color bar) for the 
          {L21b} 
          zero point (left column) and the new zero point in this paper (right column). The black lines correspond to the values of the nine ticks in the color bar.
          To visualize the $\beta$ effect, each \Zthree\ is evaluated at the ecliptic south pole (top panels), equator (middle panels), and north pole (bottom panels).
          I note that for $\GG < 12.9$ \Zthree\ has no color dependence in the $\nueff = 1.1-1.24$~\microninv\ range, as $q_{30} = 0$. Compare with the top panel in Fig.~20 of 
          {L21b}. 
          }
 \label{zpplot}
\end{figure*}

\subsection{What was done with the sample}          

$\,\!$\indent For the stars in the 26 open clusters, six globular clusters, the LMC, and the SMC I used   
{Eqn.~\ref{pic} with the \Zthree\ from L21b ($Z_{\rm EDR3,Lin}$) to calculate the group parallaxes and 
applied Eqn.~\ref{Deltapi} to obtain $\Delta\varpi_{\rm Lin}$ for each star.} 
I note that in the case of the LMC and SMC I use the measured group parallax as a reference, not the expected value from external measurements, 
which is within one sigma of the uncertainty (including the angular covariance terms) but off by a few \microas\ (Paper I). In this way, the analysis in this paper deals with the
behavior of \Zthree\ as a function of magnitude, color, and ecliptic latitude but does not change the global anchoring of the parallaxes with respect to the QSO values or studies the
effect of the angular covariance for small or intermediate angles.

As previously mentioned, when normalized by its uncertainty, $\Delta\varpi$ should have a mean of zero and a standard deviation of one. As discussed in the next section, I detected that 
is not the case for the combination of $\Delta\varpi_{\rm Lin}$ and the internal uncertainties: the standard deviation is larger than one for all magnitudes and the mean is very close
to one for $\GG > 13$ but deviates in some cases for brighter stars. The first effect means that $k$ is significantly different from one (even after accounting for the effect of \sigmas\
in Eqn.~\ref{sigmae}) and the second one that it is possible to improve $Z_{\rm EDR3,Lin}$ for bright stars. Both of those effects were anticipated in Paper I and are corroborated here.
Therefore, after the test using $Z_{\rm EDR3,Lin}$ I derive and test an alternative zero point in the next section.

\section{Results}          

$\,\!$\indent In this section I first describe the behavior of $\Delta\varpi_{\rm Lin}$ as a function of magnitude, color, and ecliptic latitude in terms of its average, 
$\overline{\Delta\varpi}$, and (non-normalized) standard deviation, $\sigma_{\Delta\varpi}$ for the 
{L21b} 
zero point. I then explain how the new zero point is calculated. Finally, 
I compare the two results.

\subsection{The Lindegren zero point}          

$\,\!$\indent One problem with analyzing data that depends on three parameters (in this case, \GG, \nueff, and $\beta$) is how to display them. Given the characteristics of the data, 
I do it in in multiple ways\footnote{In some cases, the figures/tables show/list the information relevant only to the 
{L21b}, 
others combine it with the equivalent for the zero point in this paper.}: 

\begin{enumerate}
 \item Figure~\ref{gdpi_orig} plots $\Delta\varpi_{\rm Lin}$ as a function of \GG\ for the whole sample, as that is the parameter that most influences the quality of the parallaxes (to 
       first order, the internal uncertainty is a function of \GG). In that figure I also show $\overline{\Delta\varpi}$, $\sigma_{\Delta\varpi}$, and the average (internal) parallax uncertainty 
       as a function of \GG\ using error bars (left panel) and lines (right panel). 
 \item The green line in Fig.~\ref{k} plots the $k$ value derived from the results plotted in Fig.~\ref{gdpi_orig} using Eqn.~\ref{sigmae} and assuming that $\sigma_{\Delta\varpi}$ is the 
       external uncertainty.
 \item The left panel of Fig.~\ref{dpihisto} displays a 2-D histogram of $\overline{\Delta\varpi}$ as a function of \GG\ and \nueff\ (i.e., a CMD) and in Table~\ref{stats_orig} the values of a 
       simplified version of the histogram together with the values of $\sigma_{\Delta\varpi}$ and $k$ are given. 
 \item The inclusion of the effect of the ecliptic latitude is more difficult to visualize because most of the sample is either from the LMC (80.8\%), located around the south ecliptic pole, or 
       the SMC (16.3\%), not too far from it. On the other hand, the remaining sample (clusters, 2.9\%) is more uniformly distributed over the celestial sphere (Fig.~\ref{lathisto}). Therefore, to 
       compare the effect between a sample that changes litlle in $\beta$ with one that is more distributed, I use Figs.~\ref{dpihisto}~and~\ref{dpihistoclus}, keeping in mind Fig.~\ref{CMDs} to 
       remember in which areas of the CMD the LMC and SMC population are not dominant. 
 \item To visualize the fitted functions themselves, I use Fig.~\ref{zpplot}, which is inspired on Fig.~20 of 
       {L21b} 
       but with two differences: the full functions are plotted (as opposed to 
       individual points) and three panels are used for each fit to show the changes induced by $\beta$. In that respect, I note that $q_{00}$, $q_{01}$, and $q_{02}$ shift the different \GG\ 
       sections up and down between panels but do not change the overall aspect of the plot. That is caused by $q_{11}$ the mixed color-latitude term and that is why the largest difference between 
       the three left plots in Fig.~\ref{zpplot} takes place around $\GG=12$, as that is the magnitude at which the $q_{11}$ in $Z_{\rm EDR3,Lin}$ is larger in the $\GG=6-18$~range.
\end{enumerate}

I analyze the 
{L21b} 
zero point using the same three magnitude ranges previously described: faint ($13 < \GG < 18$), intermediate ($9.2 < \GG < 13$), and bright ($6 < \GG < 9.2$).

\subsubsection{Faint range: $13 < \GG < 18$}

$\,\!$\indent With some minor exceptions, the 
{L21b} 
zero point works very well in the faint magnitude range. $\overline{\Delta\varpi}$ stays close to zero for all magnitudes in Fig.~\ref{gdpi_orig} 
and that is reflected in the first relevant column of Table~\ref{stats_orig}, where all absolute values of $\overline{\Delta\varpi}$ are less than 1~\microas\ for $\GG > 13.5$ and $\sim$2~\microas\ 
in the $\GG=13.0-13.5$~range. Equivalently, mosts cells in the left panel of Fig.~\ref{dpihisto} for $\GG > 13$. The only significant local effect in Fig.~\ref{gdpi_orig} is caused by the 
concentration of sources around $\GG = 13.8$ (red clump stars in the relatively metal-rich globular cluster 47~Tuc), but that is just an effect of a few \microas. In the left panels of 
Figs~\ref{dpihisto}~and~\ref{dpihistoclus} (especially in the second case), I see a larger effect: for both very blue stars (mostly extreme horizontal branch stars in globular clusters) and very 
red stars with $\GG > 16$, $\overline{\Delta\varpi}$ is $< -40$~\microas\ in Fig.~\ref{dpihistoclus} and shows negative values in Fig.~\ref{dpihisto}, with the caveat that the number of stars per
cell is small.  
{That is the one of the few CMD regions where the L21b} 
zero point might be improved in the faint range. 

\subsubsection{Intermediate range: $9.2 < \GG < 13$}

$\,\!$\indent The situation is different in the intermediate range. For $11 < \GG < 13$, $\overline{\Delta\varpi}$ is consistently negative in Fig.~\ref{gdpi_orig} and in
{the first group of columns in Table~\ref{stats_orig}, } 
with indication of substructures as a function of \GG\ at least for $12 < \GG < 13$ and possibly also for $11 < \GG < 12$. The two additional $\overline{\Delta\varpi}$ 
columns in Table~\ref{stats_orig} and the left panel of Fig.~\ref{dpihisto} reveal that the deviations are significantly larger for blue stars, with a value of almost $-17$~\microas\ for the 242 
stars with $11.5 < \GG < 12.0$ and $\nueff > 1.5$~\microninv. This effect was already hinted at by 
{L21b} 
(their subsection 6.2) in the LMC data but could not be better checked by them due to 
the absence of a larger sample. For $9.2 < \GG < 11$, $\overline{\Delta\varpi}$ is also negative for blue stars (but by a smaller amount than for $11 < \GG < 12$) but for red stars it is positive 
(with a smaller sample). Therefore, the 
{L21b} 
zero point can be significantly improved in this range.

\subsubsection{Bright range: $6 < \GG < 9.2$}

$\,\!$\indent In this range the number of stars is smaller, so the analysis of the 
{L21b} 
zero point cannot be as thorough. Two main issues can be described. 
First, $\sigma_{\Delta\varpi}$ increases significantly, with a large effect on $k$ (see below). Second, $\overline{\Delta\varpi}$ has the largest deviations from zero of all ranges, following a
sequence of negative-positive-negative values as 
{one progresses} 
toward brighter magnitudes. However, those aspects have to be qualified by the small sample and by almost all bright stars in the sample being blue. It is possible to improve the 
{L21b} 
zero point in this range but being subject to larger uncertainties in the outcome.

\begin{table*}
\caption{$\Delta Z_{\rm EDR3}(G,\nueff,\beta)$ coefficients for 5-parameter solutions calculated in this paper.
         The last two columns show the combinations of the first five coefficients evaluated at the south ecliptic pole.}
\centerline{
\begin{tabular}{rrrrrrrrrrr}
\hline
\mci{\GG} & $q_{00}$  & $q_{01}$  & $q_{02}$  & $q_{10}$  & $q_{11}$  & $q_{20}$  & $q_{30}$  & $q_{40}$  & $q_{00}-q_{01}$      & $q_{10}-q_{11}$ \\
          &           &           &           &           &           &           &           &           & $+\frac{2}{3}q_{02}$ &                 \\
\hline
      6.0 & $ -27.35$ & $  -2.35$ & $  -2.01$ & $   -5.9$ & $  +19.1$ & $  -1272$ &       --- & $ -358.1$ & $ -26.34$ & $  -25.0$ \\
      7.4 & $ +18.88$ & $  -2.35$ & $  -2.01$ & $   -5.9$ & $  +19.1$ & $  -1272$ &       --- & $ -358.1$ & $ +19.89$ & $  -25.0$ \\
      9.2 & $  +0.04$ & $  -2.35$ & $  -2.01$ & $   -5.9$ & $  +19.1$ & $  -1272$ &       --- & $ -358.1$ & $  +1.05$ & $  -25.0$ \\
     10.8 & $  +7.20$ & $  +0.29$ & $ -12.56$ & $  -23.7$ & $  +19.1$ & $  -1272$ &       --- & $  -78.6$ & $  -1.46$ & $  -42.8$ \\
     11.2 & $  -3.05$ & $  -2.98$ & $  -3.57$ & $  -43.9$ & $   +1.6$ & $  -1272$ &       --- & $ +203.1$ & $  -2.45$ & $  -45.5$ \\
     11.8 & $  -3.41$ & $  -1.47$ & $ -14.93$ & $  -32.0$ & $   +1.6$ & $  -1272$ &       --- & $ -155.3$ & $ -11.89$ & $  -33.6$ \\
     12.2 & $  -3.34$ & $  -3.60$ & $  +6.46$ & $   +1.5$ & $  +22.6$ & $  -2368$ &       --- & $ -144.2$ & $  +4.57$ & $  -21.1$ \\
     12.9 & $  -6.96$ & $  -8.99$ & $  +5.91$ & $   -9.2$ & $  +22.6$ & $  -3096$ &       --- & $  +23.6$ & $  +5.97$ & $  -31.8$ \\
     13.1 & $  +0.48$ & $  +0.70$ & $  +4.27$ & $   +2.8$ & $  +15.6$ & $   -418$ & $  -75.8$ & $   -4.8$ & $  +2.63$ & $  -12.8$ \\
     15.9 & $  +5.51$ & $  -0.46$ & $  -8.92$ & $   -9.1$ & $   -6.5$ & $   -152$ & $  -75.8$ & $  -25.9$ & $  +0.02$ & $   -2.6$ \\
     16.1 & $  -2.13$ & $  -4.14$ & $  -3.11$ & $  -56.8$ & $  -53.6$ & $   -301$ & $   +6.6$ & $  -17.6$ & $  -0.06$ & $   -3.2$ \\
     17.5 & $  +4.16$ & $  +5.92$ & $  +4.16$ & $  -39.2$ & $  -53.6$ & $   -119$ & $   +6.6$ & $ -107.0$ & $  +1.01$ & $  +14.4$ \\
     18.0 & $  +2.71$ & $  -1.53$ & $  -4.55$ & $  -46.5$ & $  -53.6$ & $   -119$ &       --- & $ -118.7$ & $  +1.21$ & $   +7.1$ \\
\hline
\end{tabular}
}
\label{DeltaZ}
\end{table*}

\begin{table*}
\caption{$Z_{\rm EDR3,new}(G,\nueff,\beta)$ coefficients for 5-parameter solutions calculated from L21b and the results in this paper.}
\centerline{
\begin{tabular}{rrrrrrrrr}
\hline
\mci{\GG} & $q_{00}$  & $q_{01}$  & $q_{02}$  & $q_{10}$  & $q_{11}$  & $q_{20}$  & $q_{30}$  & $q_{40}$  \\
\hline
      6.0 & $ -54.33$ & $ -11.97$ & $ +25.39$ & $  -31.0$ & $  +19.1$ & $  -2529$ &       --- & $ -358.1$ \\
      7.4 & $  -8.17$ & $ -10.06$ & $ +24.12$ & $  -13.4$ & $  +23.7$ & $  -2529$ &       --- & $ -358.1$ \\
      9.2 & $ -27.11$ & $  -7.60$ & $ +22.48$ & $   +9.3$ & $  +29.6$ & $  -2529$ &       --- & $ -358.1$ \\
     10.8 & $ -20.03$ & $  -2.78$ & $ +10.48$ & $  +11.6$ & $  +34.8$ & $  -2529$ &       --- & $  -78.6$ \\
     11.2 & $ -33.38$ & $ -12.21$ & $  +5.51$ & $ -132.3$ & $  -10.2$ & $  -2529$ &       --- & $ +203.1$ \\
     11.8 & $ -36.95$ & $ -11.55$ & $  -1.65$ & $ -158.7$ & $  +13.2$ & $  -2529$ &       --- & $ -155.3$ \\
     12.2 & $ -16.99$ & $  -3.67$ & $ +15.81$ & $ -109.9$ & $  +63.2$ & $  -3625$ &       --- & $ -144.2$ \\
     12.9 & $ -26.49$ & $ -10.63$ & $ +21.77$ & $  -76.0$ & $  +43.2$ & $  -4353$ &       --- & $  +23.6$ \\
     13.1 & $ -37.51$ & $  +3.33$ & $ +20.41$ & $   -2.9$ & $  +29.6$ & $  -1675$ & $  +32.1$ & $  +99.5$ \\
     15.9 & $ -32.82$ & $  +5.15$ & $  +6.50$ & $   -9.1$ & $  +12.2$ & $  -1341$ & $ +168.0$ & $ +129.3$ \\
     16.1 & $ -33.18$ & $  -1.31$ & $  +5.48$ & $  -56.8$ & $  -38.1$ & $  -1705$ & $ +112.1$ & $ +153.1$ \\
     17.5 & $ -25.02$ & $  +5.83$ & $  +6.57$ & $  -39.2$ & $  -29.1$ & $  -1284$ & $ +196.3$ & $ +218.0$ \\
     18.0 & $ -22.88$ & $  +0.40$ & $  -5.10$ & $  -46.5$ & $  -35.4$ & $   -896$ & $ +126.5$ & $ +190.2$ \\
     19.0 & $ -18.40$ & $  +5.98$ & $  -6.46$ &       --- & $   +5.5$ &       --- &       --- & $ +276.6$ \\
     20.0 & $ -12.65$ & $  -4.57$ & $  -7.46$ &       --- & $  +97.9$ &       --- &       --- &       --- \\
     21.0 & $ -18.22$ & $ -15.24$ & $ -18.54$ &       --- & $ +128.2$ &       --- &       --- &       --- \\
\hline
\end{tabular}
}
\label{Znew}
\end{table*}

\subsubsection{The $k$ multiplicative constant}

$\,\!$\indent The behavior of $k$ in Fig.~\ref{k} for 
{L21b} 
has a qualitative similar one to the approximation derived in Paper~I for $\GG > 10$. An increase from $\sim 1.1$ at 
$\GG = 18$ as we move toward $\GG \sim 12$ followed by a decrease as we move toward $\GG = 10$. However, some (relatively small) differences are seen in that range, not surprising 
given the better sampling of the data here: the peak around $\GG\sim 12$ is taller, wider, and has some structure. The situation is very different for $\GG < 10$, a range that was not 
probed in Paper~I. Starting around $G=10$, where $k\sim 1.3$, it grows to values close to 2.0 around $\GG=9$ and close to 3.0 for the brightest stars sampled here. Pending the 
derivation of the new zero point below, I defer until then the analysis of the consequences of this effect.

\subsection{Calculating the new zero point}          

\begin{figure*}
 \centerline{\includegraphics[width=0.50\linewidth]{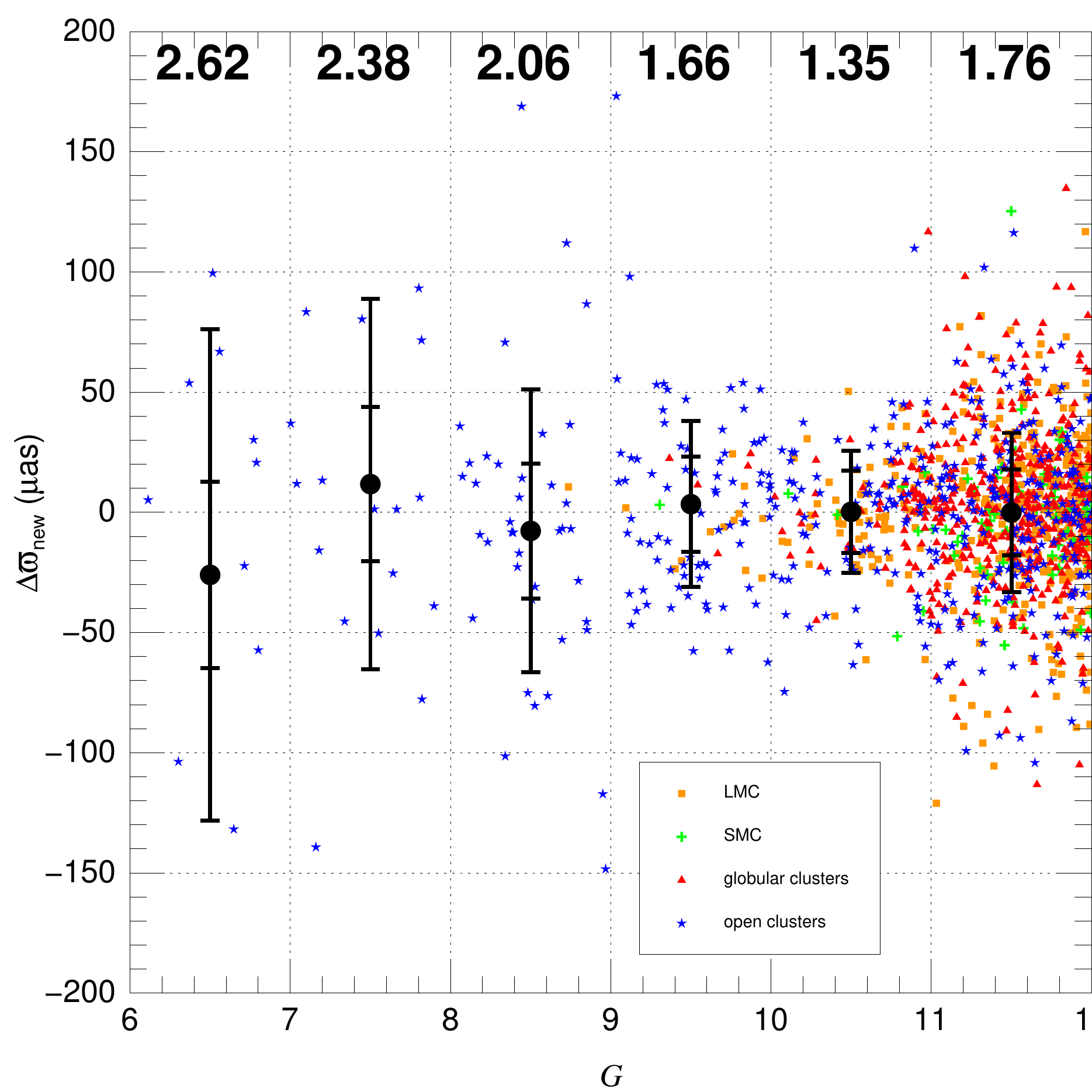}\includegraphics[width=0.50\linewidth]{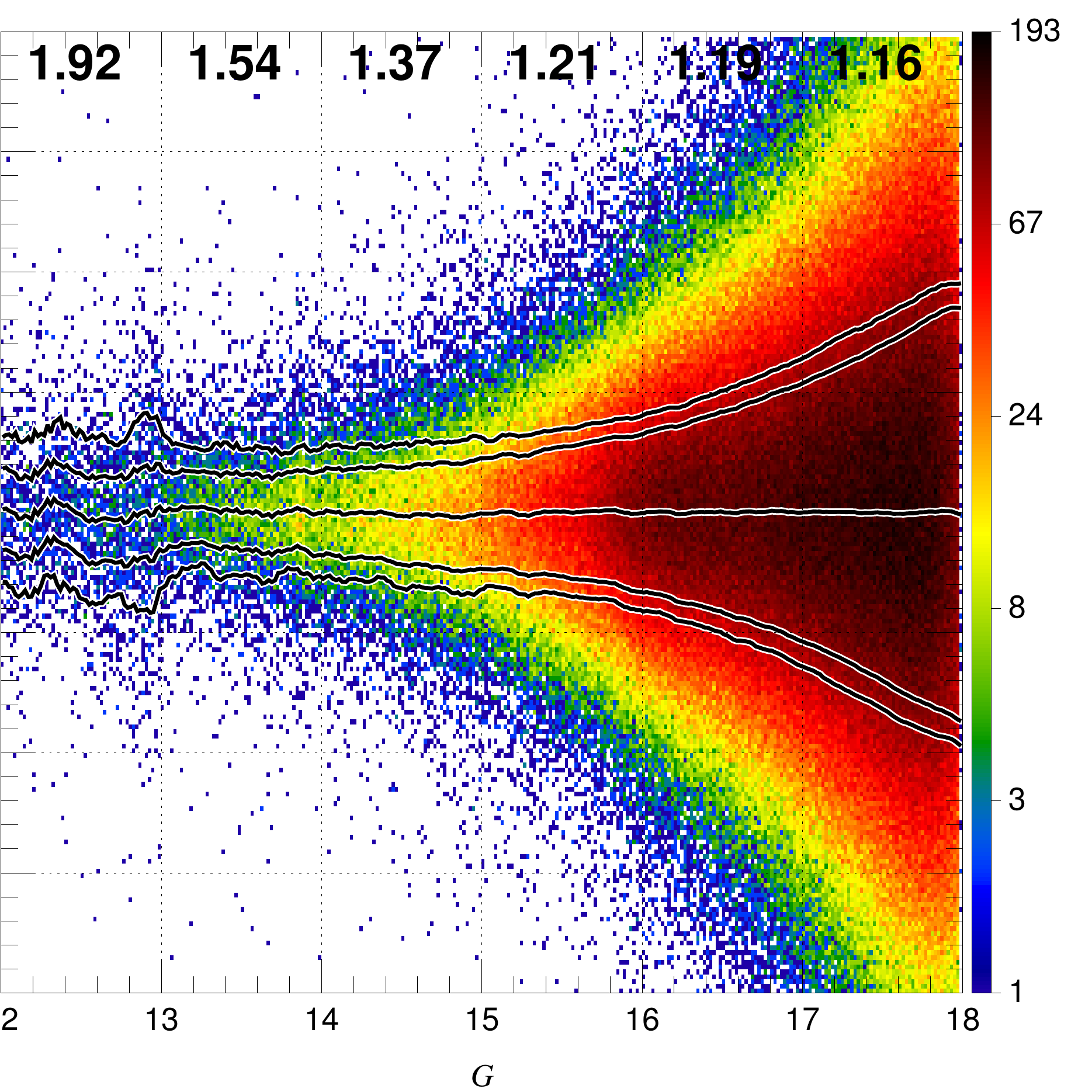}}
 \caption{Same as Fig.~\ref{gdpi_orig}, but using the zero point proposed here.}
 \label{gdpi}
\end{figure*}

$\,\!$\indent As just described, the Lindegren zero point works reasonably well for most values of its parameters, \GG, \nueff, and $\beta$, but can be tweaked in some
circumstances. Based on that, I define the new zero point as the sum of the Lindegren one and a correction term:

\begin{equation}
Z_{\rm EDR3,new} = Z_{\rm EDR3,Lin} + \Delta Z_{\rm EDR3}.
\label{Ztotdef}
\end{equation}

The goal is to derive the optimal $\Delta Z_{\rm EDR3}$ 
{by fitting $\Delta\varpi_{\rm Lin}$} 
to obtain $Z_{\rm EDR3,new}$. This strategy works because the zero point is the sum of a series of linear terms. The decomposition 
into two terms has several advantages. First, it allows us to characterize the impact of the correction better. Second, I can fit just a selection of the whole set of coefficients,
leaving the ones from 
{L21b} 
that do not require changes to be left in place. And third, it allows the anchoring of the parallaxes to remain unchanged, as previously mentioned.

To calculate $\Delta Z_{\rm EDR3}$ I assumed the same functional form as 
{L21b} 
and added three magnitude breakpoints (knots) at \GG~=~7.4,~9.2,~and~18.0. The last one is added because
it is the end of the magnitude range in our sample. The first two are added based on Fig.~\ref{gdpi_orig}, as the behavior of $\Delta\varpi$ in \GG\ there is not linear in the 
\GG~=~6.0-10.8 range and those are the apparent magnitudes at which the behavior changes (but see below for the restrictions placed on the fitted coefficients in that range).

To fit $\Delta Z_{\rm EDR3}$ I wrote a program in IDL based on the MPFIT package 
\citep{Mark09}\footnote{Available from \url{http://purl.com/net/mpfit}.}. MPFIT allows for arbitrary 
functions to be fitted to almost any type of data while fixing the value of some coefficients and tying up others among them. As 
{L21b} 
correctly cautions, it is important to avoid
overfitting. Therefore, initially I only fitted the coefficients where enough data were present and, by trial and error, I added additional restrictions. In the end, the following
coefficients for $\Delta Z_{\rm EDR3}$ were fit:

\begin{itemize}
 \item For the three breakpoints at \GG~=~6.0,~7.4,~and~9.2, only $q_{00}$ was fit at each one of them. $q_{30}$ was not fit (as there are no stars as red as needed, see below) and 
       the rest of the coefficients were tied up, that is, forced to have the same values at the three breakpoints. For $q_{11}$ the range of bright magnitudes where the coefficients 
       are tied up was extended to \GG~=~10.8 and for $q_{20}$ to \GG~=~10.8,~11.2,~and~11.8.
 \item The mixed latitude-color term, $q_{11}$, was also tied up in three additional regions: (a) \GG~=~11.2~and~11.8, (b) \GG~=~12.2~and~12.9, and 
       (c) \GG~=~16.1,~17.5,~and~18.0.
 \item The cubic term for red stars, $q_{20}$ was tied up for $\GG$ at 17.5 and 18.0.
 \item As there are few very red stars in the sample, $q_{30}$ was only fit in the \GG~=13.1-17.5~range and even there, the two values at 13.1 and 15.9 were tied up and the same was
       done for the two values at 16.1~and~17.5. 
\end{itemize}

The fitted coefficents for $\Delta Z_{\rm EDR3}$ are given in Table~\ref{DeltaZ}. The last two columns give the combinations of the first five coefficients at the south ecliptic pole,
which is a reasonable approximation for the LMC stars. The final $Z_{\rm EDR3,new}$ is given in Table~\ref{Znew}, where for magnitudes fainter than 
\GG~=~18.0 the values given are simply those of $Z_{\rm EDR3,Lin}$, as the sample in this paper does not reach those very faint magnitudes. The values of the new zero point are plotted
in the right column of Fig.~\ref{zpplot}, allowing for a direct comparison with the 
{L21b} 
zero point. To facilitate the implementation of the new results, two IDL routines are given in
Tables~\ref{spicor}~and~\ref{zpedr3}. The first one produces the $k$ multiplicative constant and the second one the new zero point.

{As a final note, I am fitting $\Delta\varpi_{\rm Lin}$ to obtain a new $\Delta Z_{\rm EDR3}$ and the values of $\Delta\varpi_{\rm Lin}$ depend on the group parallaxes 
calculated with the L21b zero point. In principle, one could iterate the process by using the new group parallaxes obtained with $Z_{\rm EDR3,new}$ (i.e., the ones that are listed in 
Villafranca~II for the open clusters) and calculate a new incremental zero point to be added to the previous one. However, that was found unnecesary because the differences between
the old and new group parallaxes are small ($\sim 1$~\microas) and unbiased (some positive and some negative), leading to very similar values of $Z_{\rm EDR3,new}$.} 

\subsection{Analysis}

\begin{table*}
\caption{Statistics (in \microas) by \GG\ magnitude and \nueff\ ranges using the results in this paper.}
\centerline{
\begin{tabular}{crrrrrrrrrrrr}
\hline
          & \mciv{all} & \mciv{$\nueff > 1.5$} & \mciv{$\nueff < 1.5$} \\
\GG       & $N$    & $\overline{\Delta\varpi}_{\rm new}$ & $\sigma_{\Delta\varpi,{\rm new}}$ & \mci{$k_{\rm new}$} & $N$    & $\overline{\Delta\varpi}_{\rm new}$ & $\sigma_{\Delta\varpi,{\rm new}}$ & \mci{$k_{\rm new}$} & $N$    & $\overline{\Delta\varpi}_{\rm new}$ & $\sigma_{\Delta\varpi,{\rm new}}$ & \mci{$k_{\rm new}$} \\
\hline
  6.0- 7.0 & \num{11}     & $ -26.01$ & $ 102.14$ & 2.62 &          --- &       --- &       --- &  --- &          --- &       --- &       --- &  --- \\
  7.0- 8.0 & \num{18}     & $ +11.76$ & $  77.00$ & 2.38 &          --- &       --- &       --- &  --- &          --- &       --- &       --- &  --- \\
  8.0- 9.0 & \num{40}     & $  -7.72$ & $  58.77$ & 2.06 &          --- &       --- &       --- &  --- &          --- &       --- &       --- &  --- \\
  9.0-10.0 & \num{98}     & $  +3.41$ & $  34.52$ & 1.66 & \num{84}     & $  +3.72$ & $  36.64$ & 1.70 & \num{14}     & $  +1.53$ & $  17.79$ & 1.01 \\
 10.0-10.5 & \num{154}    & $  -2.14$ & $  23.02$ & 1.22 & \num{113}    & $  -1.31$ & $  24.51$ & 1.27 & \num{41}     & $  -4.43$ & $  18.39$ & 1.03 \\
 10.5-11.0 & \num{170}    & $  +1.80$ & $  26.08$ & 1.41 & \num{94}     & $  +0.48$ & $  27.20$ & 1.48 & \num{76}     & $  +3.43$ & $  24.71$ & 1.31 \\
 11.0-11.5 & \num{340}    & $  +0.07$ & $  35.55$ & 1.76 & \num{142}    & $  -1.49$ & $  41.03$ & 2.02 & \num{198}    & $  +1.19$ & $  31.09$ & 1.54 \\
 11.5-12.0 & \num{637}    & $  -0.17$ & $  31.79$ & 1.76 & \num{242}    & $  -2.68$ & $  35.03$ & 1.83 & \num{395}    & $  +1.37$ & $  29.58$ & 1.69 \\
 12.0-12.5 & \num{1098}   & $  +1.27$ & $  32.82$ & 1.90 & \num{394}    & $  +2.57$ & $  37.36$ & 1.85 & \num{704}    & $  +0.55$ & $  29.98$ & 1.91 \\
 12.5-13.0 & \num{1768}   & $  -1.92$ & $  36.63$ & 1.92 & \num{758}    & $  -2.13$ & $  38.39$ & 1.78 & \num{1010}   & $  -1.77$ & $  35.26$ & 2.06 \\
 13.0-13.5 & \num{2712}   & $  +1.31$ & $  26.76$ & 1.61 & \num{1380}   & $  +3.29$ & $  25.11$ & 1.44 & \num{1332}   & $  -0.73$ & $  28.23$ & 1.80 \\
 13.5-14.0 & \num{4877}   & $  +0.05$ & $  26.34$ & 1.50 & \num{2886}   & $  +0.62$ & $  26.05$ & 1.44 & \num{1991}   & $  -0.78$ & $  26.75$ & 1.59 \\
 14.0-14.5 & \num{7607}   & $  -0.28$ & $  28.17$ & 1.39 & \num{4060}   & $  -0.77$ & $  28.10$ & 1.34 & \num{3547}   & $  +0.27$ & $  28.23$ & 1.45 \\
 14.5-15.0 & \num{14602}  & $  -1.37$ & $  31.31$ & 1.35 & \num{7716}   & $  -1.56$ & $  33.76$ & 1.40 & \num{6886}   & $  -1.16$ & $  28.32$ & 1.28 \\
 15.0-15.5 & \num{32994}  & $  -0.22$ & $  32.70$ & 1.24 & \num{11952}  & $  +0.53$ & $  36.87$ & 1.33 & \num{21042}  & $  -0.64$ & $  30.08$ & 1.17 \\
 15.5-16.0 & \num{78681}  & $  +0.15$ & $  37.67$ & 1.20 & \num{19235}  & $  +0.71$ & $  43.98$ & 1.34 & \num{59446}  & $  -0.04$ & $  35.38$ & 1.15 \\
 16.0-16.5 & \num{146663} & $  -0.06$ & $  45.40$ & 1.19 & \num{32112}  & $  -0.15$ & $  52.77$ & 1.31 & \num{114551} & $  -0.04$ & $  43.11$ & 1.14 \\
 16.5-17.0 & \num{206874} & $  +0.17$ & $  57.48$ & 1.18 & \num{54569}  & $  +0.38$ & $  65.58$ & 1.30 & \num{152305} & $  +0.09$ & $  54.29$ & 1.13 \\
 17.0-17.5 & \num{303660} & $  +0.01$ & $  73.62$ & 1.18 & \num{97619}  & $  +0.16$ & $  81.93$ & 1.29 & \num{206041} & $  -0.06$ & $  69.34$ & 1.12 \\
 17.5-18.0 & \num{373782} & $  -0.09$ & $  89.77$ & 1.14 & \num{132270} & $  +0.54$ & $  96.27$ & 1.21 & \num{241512} & $  -0.44$ & $  86.00$ & 1.09 \\
\hline
\end{tabular}
}
\label{stats}
\end{table*}

\begin{figure*}
 \centerline{\includegraphics[width=0.49\linewidth]{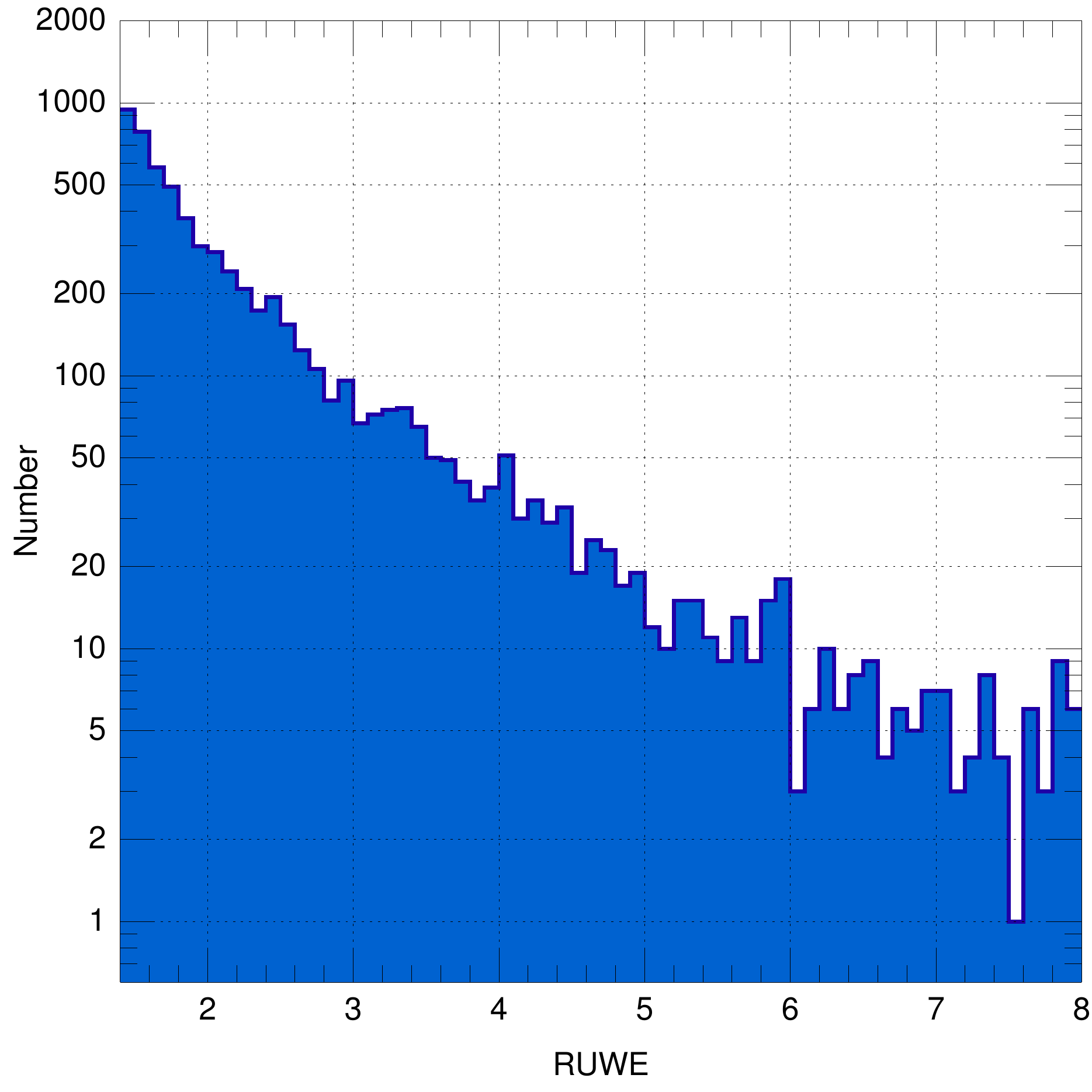}\
             \includegraphics[width=0.49\linewidth]{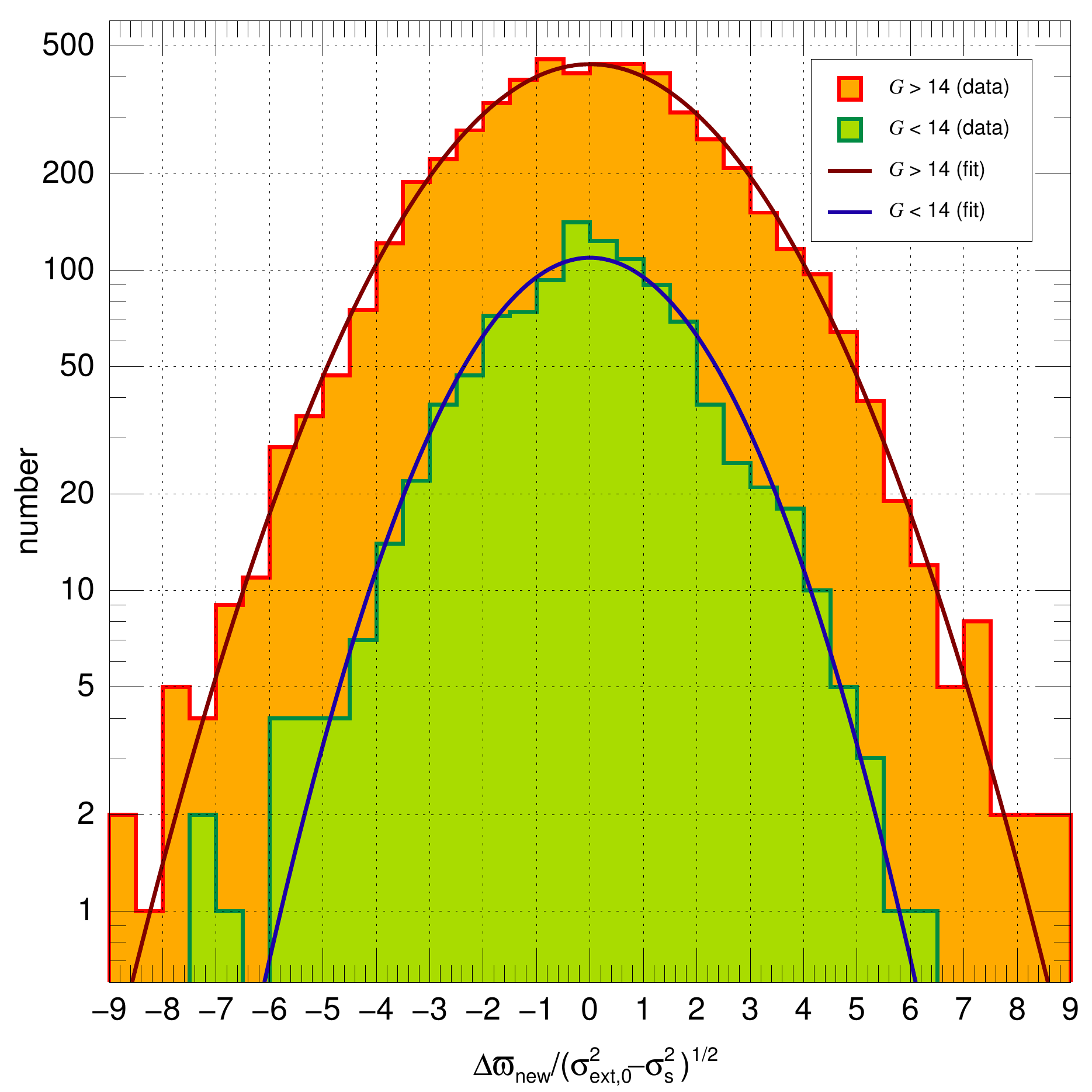}}
 \caption{(left) RUWE histogram of five-parameter sources from the 32 stellar clusters in our sample used to evaluate $k_{\rm ext}$.
          (right) Normalized parallax histograms for faint and bright sources with RUWE between 1.4 and 8.0. The two fits correspond to values of $k_{\rm ext}$ of 2.36 and 1.88, respectively.}
 \label{kextra}
\end{figure*}

$\,\!$\indent I now analyze $Z_{\rm EDR3,new}$ by 
{using it to calculate the new group parallaxes and} 
studying the residuals using the new zero point, $\Delta\varpi_{\rm new}$. For that purpose I use
Fig.~\ref{gdpi}, the equivalent to Fig.~\ref{gdpi_orig}, Table~\ref{stats}, the equivalent to Table~\ref{stats_orig}, and the already introduced 
Figs.~\ref{k},~\ref{dpihisto},~\ref{dpihistoclus},~and~\ref{zpplot} and Tables~\ref{DeltaZ}~and~\ref{Znew}. As I did for the Lindegren zero point, I divide the analysis in the same three 
magnitude ranges.

\subsubsection{Faint range: $13 < \GG < 18$}

$\,\!$\indent As expected, the differences between the Lindegren and the new zero points are small for most values of the parameter space for faint stars. This is especially true near the south
ecliptic pole (in our sample, the LMC and SMC), as evidenced by the small values in the last two columns of Table~\ref{DeltaZ} and the similar appearance of the top two panels in Fig.~\ref{zpplot} for 
$\GG > 13$. However, in some cases the differences are significant:

\begin{itemize}
 \item The largest one is the existence of a relatively large $\Delta Z_{\rm EDR3}$ mixed color-latitude term ($q_{11}$) for $\GG > 16.1$. While the top two plots of Fig.~\ref{zpplot} are similar for 
       those magnitudes, the bottom two plots are quite different. 
 \item The new zero point also has a significant $q_{40}$ term (blue stars) in this magnitude range, as established with the help of extreme horizontal branch stars in globular clusters. 
       $\Delta\varpi_{\rm new}$ is smaller than $\Delta\varpi_{\rm Lin}$ for them (bottom left corner of both panels in Fig.~\ref{dpihistoclus}, the effect is also seen to some degree in 
       Fig.~\ref{dpihisto}). However, some residuals are still seen, especially for stars bluer than $\nueff = 1.8$~\microninv. At this stage it is not clear whether the effect is caused by small 
       number statistics or by the need of using a different functional form for the zero point
       {(e.g., a $q_{41}$ term),} 
       but I note that the effect is also seen in the bottom two panels of Fig.~10 in L21b.
 \item {L21b} 
       detected the existence of a ``hook'' in the behavior of the reddest stars ($\nueff < 1.18$~\microninv) for $\GG > 16$. The effect is also seen in the bottom right corners of 
       Figs.~\ref{dpihisto}~and~\ref{dpihistoclus} but as the functional form does not include a term there, the new zero point does not correct for it and the appearance does not change significantly
       between the left and right panels.
\end{itemize}

Nevertheless, we should bear in mind that for most stars in this magnitude range the effect is small. In that way, Figs~\ref{gdpi_orig}~and~\ref{gdpi} are very similar for $\GG > 13$. Doing a more
quantitative comparison, nine out of the thirty values of $\overline{\Delta\varpi}$ in Table~\ref{stats_orig} have absolute values larger than 1~\microas, a number that in Table~\ref{stats} is reduced to five
(very good statistics if one bears in mind that $\sigmas = 10.3$~\microas). In summary, for $13 < \GG < 16$ the new zero point is very similar to the Lindegren zero point and for $16 < \GG < 18$ some 
differences appear for stars far from the south ecliptic pole and for very blue and very red stars.

\subsubsection{Intermediate range: $9.2 < \GG < 13$}

$\,\!$\indent The new zero point is significantly different from the Lindegren one for intermediate magnitudes, even for positions close to the south ecliptic pole (Fig.~\ref{zpplot}). For 
$\beta=-90^{\rm o}$, $Z_{\rm EDR3,new}$ is qualitatively similar to $Z_{\rm EDR3,Lin}$ (with a reversal of the behavior as a function of color with respect to fainter magnitudes in both cases) but 
quantitatively different for $11 < \GG < 13$. For brighter stars (continuing to $\GG = 6$) a large $q_{40}$ term appears, making $Z_{\rm EDR3,new}$ more negative for blue stars. As for the changes as 
a function of $\beta$, they are small for $11 < \GG < 12$ and significant otherwise. 

Analyzing the $\overline{\Delta\varpi}$ values in Table~\ref{stats_orig} for $9 < \GG < 13$ I find just six out of twenty one with absolute values smaller than 2~\microas\ and five with absolute values
larger than 5~\microas. On the other hand, in 
{Table~\ref{stats}} 
thirteen out of twenty one values have absolute values smaller than 2~\microas\ and there are none above 5~\microas. It is clear that 
the new zero point lowers the residuals and provides a better fit. The effect is also seen in the comparison between Figs~\ref{gdpi_orig}~and~\ref{gdpi}, especially for $11 < \GG < 13$. I note the
apparent existence of some fine structure as a function of \GG\ with an amplitude of a few \microas\ for at least $12 < \GG < 13$ that could be further corrected in the future by introducing additional 
magnitude breakpoints.

The improvement is not the same for all magnitudes and colors. From Tables~\ref{stats_orig}~and~~\ref{stats} and Figs.~\ref{dpihisto}~and~\ref{dpihistoclus}, we see that it is larger for blue stars than
for red ones and for $11 < \GG < 12$ than for other magnitudes in this range. The main shortcoming of the new zero point arises from the relatively small sample size, especially for the brighter part of 
the magnitude range for red stars. One possible improvement in this sense would be to add open clusters with red supergiants to the analysis. In summary, for $9.2 < \GG < 13$ the new zero point provides a
significant improvement with respect to the Lindegren zero point.

\subsubsection{Bright range: $6 < \GG < 9.2$}

$\,\!$\indent Once we get to this magnitude range, the sample is highly incomplete and consists almost exclusively of blue stars. $\Delta Z_{\rm EDR3}$ has large $q_{00}$ and (as already mentioned)
$q_{40}$ terms, leading to significant changes in the overall zero point and on its color dependence for blue stars (Fig.~\ref{zpplot}). $Z_{\rm EDR3,new}$ is capable of partially flattening the behavior
of $\overline{\Delta\varpi}$ in Fig.~\ref{gdpi} and Table~\ref{stats} with respect to Fig.~\ref{gdpi_orig} and Table~\ref{stats_orig}. Still, the values are not zero and the most visible characteristic is
the persistence of the large values of $\sigma_{\Delta\varpi}$, which in turn lead to the same effect for $k$, as we see in the next subsection. In summary, for bright stars the new zero point
improves upon the Lindegren zero point but provides little information for red stars and this magnitude range is dominated by the effect of the large dispersion of the results, indicating a significant
underestimation of the external parallax uncertainties by the internal values.

\subsubsection{The $k$ multiplicative constant}

$\,\!$\indent Table~\ref{stats} lists the values of $k$ as a function of magnitude using the new zero point, including all colors in the calculation or just the blue or the red stars. The values are also
plotted in Fig.~\ref{k} and can be compared with the results form the Lindegren zero point in Table~\ref{stats_orig} and in the same figure. For faint stars the results for both zero points are identical,
an indication of the similarity between the two zero points in that magnitude range. For intermediate and bright stars the new zero point reduces the value of $k$ but only slightly so. This must be
interpreted as the effect of $\Delta Z_{\rm EDR3}$, the transformation from the Lindegren zero point to the new one (the correction of a systematic effect), being in general small compared to the true 
random uncertainties. The ultimate reason why we can measure $\Delta Z_{\rm EDR3}$ is because we are using large numbers of stars in each cluster or galaxy. On the other hand, the introduction of the $k$ 
value (and, to a lesser extent, of \sigmas) in the transformation of internal uncertainties to external ones is an important effect, given that it is significantly larger than one.

The comparison between blue and red stars in Table~\ref{stats} shows a similar behavior as a function of \GG, indicating that a single $k(G)$ is a good approximation. Nevertheless, there are some
differences. For most magnitudes $k$ appears to be slightly lower for red stars. The exception is the region around $\GG\approx 12.7$, where the local maximum in $k$ for red stars is located. The
equivalent maximum for blue stars is located around $\GG\approx 11.3$.

The largest values of $k$ occur for bright stars ($\GG < 9.2$) and that is the single most important conclusion of this work: using the internal uncertainties leads to a significant undestimation of their
{\it Gaia}~EDR3 distance uncertainties. The effect is also important for stars with $11 < \GG < 13$. An example of the effect is seen in the case of the eleven intermediate/bright stars listed in
Table~\ref{addedbyhand}: all of their {\it Gaia}~EDR3 parallaxes are within 3 sigmas of the group values if one uses the $k(G)$ in Table~\ref{stats}.

\subsubsection{Objects with large RUWE}

\begin{table}
\centerline{
\begin{tabular}{rrc}
\hline
\mci{\GG} & \mci{$N$} & $k_{\rm ext}$ \\
\hline
 6-13     &       276 &          1.50 \\
13-14     &       759 &          2.01 \\
14-15     &      1050 &          2.28 \\
15-16     &      1212 &          2.38 \\
16-17     &      1230 &          2.44 \\
17-18     &      1700 &          2.32 \\
\hline
\end{tabular}
}
\label{kextra_values}
\caption{$k_{\rm ext}$ calculated in magnitude bins for five-parameter solutions using the stellar cluster data for objects with RUWE between 1.4 and 8.0. The second column gives the number of objects per bin.}
\end{table}

$\,\!$\indent The results presented so far refer to objects with five-parameter solutions and ``good'' RUWE, that is, values up to 1.4. However, as shown in Paper~I, it is possible to treat objects with RUWE larger
than 1.4 by introducing an additional factor, $k_{\rm ext}$, in Eqn.~\ref{sigmae} multiplying \sigmai, that is:

\begin{equation}
\begin{array}{rcrl}
        &   & \sqrt{              k^2 \sigma_{\rm int}^2 + \sigma_{\rm s}^2}, & {\rm RUWE} < 1.4  \\
\sigmae & = &                                                                     \\
        &   & \sqrt{k_{\rm ext}^2 k^2 \sigma_{\rm int}^2 + \sigma_{\rm s}^2}, & {\rm RUWE} > 1.4. \\
\end{array}
\label{sigmae2}  
\end{equation}

\noindent Here I present an extended analysis of this issue.

I select the five-parameter solutions in each of the 32 clusters in our sample with [a] \GG\ between 6 and 18, [b] RUWE between 1.4 and 8.0, [c] no restrictions on parallax uncertainty, and [d] the rest of the 
restrictions that apply to each individual cluster. I apply $Z_{\rm EDR3,new}$ and subtract the group parallax to each individual parallax value to calculate $\Delta\varpi_{\rm new}$ and I use Eqn.~\ref{sigmae} 
to obtain $\sigma_{\rm ext,0}$, that is, the external uncertainty assuming the same $k$ as for objects with RUWE lower than 1.4. Finally, I make a 9-sigma cut in normalized parallax, the value being so high because 
we expect the real external uncertainties to be larger than for objects with good RUWE. The final sample has 6227~objects.

The distributions of RUWE and of normalized $\Delta\varpi_{\rm new}$ are shown in Fig.~\ref{kextra}. I divide the sample in a faint and a bright range (limited by $\GG = 14$) and calculate Gaussian fits with zero mean.
The results are excellent, indicating that parallaxes with large RUWE are not strongly biased toward higher or lower values and that the inclusion of a $k_{\rm ext}$ results in external uncertainties with the proper 
behavior. $k_{\rm ext}$ is significantly larger for fainter stars, leading us to divide \GG\ in bins to tabulate its behavior. Unfortunately, there are few bright stars in this sample, so the information there is limited and
a single bin for the \GG~=6-13 has to be used. The results are given in Table~\ref{kextra_values} and have been incorporated into Table~\ref{spicor}.

\subsubsection{Objects with six-parameter astrometric solutions}

$\,\!$\indent As previously mentioned, there are not enough stars with six-parameter solutions to derive a new $Z_{\rm EDR3}$ for them. However, as we have seen, the value of $k$ is relatively robust with respect to the zero
point and can be calculated even with a relatively small number of points. In the case of our cluster sample, we have enough stars to determine it for the range $13 < \GG < 18$. The values of $k$ for six-parameter solutions
there follow the same pattern as in Fig.~\ref{k} for five-parameter solutions, growing from $\GG\sim 18$ to $\GG\sim 13$ but with values that are $\sim 1.25$ times higher. Therefore, a simple approximation is to use a value of
$k_{\rm ext}$ of 1.25 for six-parameter solutions, that is:

\begin{equation}
\sigmae = \sqrt{1.25^2 k^2 \sigma_{\rm int}^2 + \sigma_{\rm s}^2} \;\; \textrm{for six-parameter solutions} \\
\label{sigmae3}  
\end{equation}

\noindent but I note that it is not tested for $\GG < 13$. That approximation has been incorporated into Table~\ref{spicor}.

\section{Summary and future work}

$\,\!$\indent In this paper I have presented a new zero point for {\it Gaia}~EDR3 parallaxes as a function of magnitude, color, and ecliptic latitude. I have used the same functional form as 
{L21b} 
(Eqn.~\ref{Zdef}) but derived the zero point using a combination of data from the LMC, the SMC, globular clusters, and open clusters. The differences between the two zero points are small for faint stars 
($\GG > 13$) but become significant for stars brighter than that, though it should be noted that for $\GG < 9.2$ the zero point is poorly defined due to the small size of the sample. As a second result, I have 
determined that the multiplicative constant $k$ that is used to convert from internal parallax uncertainties to external ones (Eqn.~\ref{sigmae}) is significantly larger than one for most stars and even 
larger than two for $\GG < 9.2$. $k$ is found to be even larger for objects with RUWE larger than 1.4 or with six-parameter solutions. Therefore, the distance uncertainties derived assuming internal parallax 
uncertainties will be, in general, underestimated.

{\it Gaia}~DR4 is stil several years in the future and this paper has not exploited all of the possibilities for improving on the zero point. The most obvious future line of work would be adding more
clusters to improve the statistics for bright stars. It would be especially interesting to include young clusters with red supergiants, as those would extend the sampling to a larger range of colors 
{(bright red stars)} 
and of ecliptic latitudes. Red giants in additional globular clusters would also help but to a lesser degree
{regarding bright red stars,} 
given that the tip of the red giant branch (TRGB) is not too bright (bottom left panel in Fig.~\ref{CMDs}) and that there are only a few nearby globular clusters. 
{However, adding globular clusters would be helpful in calibrating two regions of the CMD: RGB stars for the intermediate/faint very red region (the dominant {\it Gaia} population in some regions of 
the Galactic plane lies there in the form of high-extinction red giants) and blue horizontal branch (BHB) stars for the faint very blue region. The latter could be used to test whether the residuals for very blue
stars described in subsection 3.3.1 could be corrected with a latitude dependence.} 
Other possibilities would be testing new basis functions and magnitude breakpoints and building a sample large enough to test the zero point for six-parameter solutions. Once {\it Gaia}~DR4 becomes
available, these same clusters could be used to determine its (hopefully smaller) parallax zero points.


\begin{acknowledgements}
I acknowledge support from the Spanish Government Ministerio de Ciencia through grant PGC2018-\num{095049}-B-C22. 
This work has made use of data from the European Space Agency (ESA) mission {\it Gaia}\footnote{\url{https://www.cosmos.esa.int/gaia}}, 
processed by the {\it Gaia} Data Processing and Analysis Consortium (DPAC\footnote{\url{https://www.cosmos.esa.int/web/gaia/dpac/consortium}}).
Funding for the DPAC has been provided by national institutions, in particular the institutions participating in the {\it Gaia} 
Multilateral Agreement. 
The {\it Gaia} data is processed with the computer resources at Mare Nostrum and the technical support provided by BSC-CNS.
\end{acknowledgements}

%
%

\bibliographystyle{aa} 
\bibliography{general} 

\vfill

\eject

\appendix

\section{IDL codes}

\begin{table}[h!]
\caption{Function that calculates {\it Gaia}~EDR3 parallax external uncertainties.}
{\scriptsize
\begin{verbatim}
FUNCTION SPICOR, spi, gmag, ruwe, npar
;;;;;;;;;;;;;;;;;;;;;;;;;;;;;;;;;;;;;;;;;;;;;;;;;;;;;;;;;;;;;;;;;;;;;;;
;                                                                     ;
; This function returns the parallax external uncertainties for Gaia  ;
;  EDR3 in milliarcseconds for an array of values. spi, gmag, ruwe,   ;
;  and npar must be arrays with the same number of elements.          ;
;                                                                     ;
; Positional parameters:                                              ;
; spi:      Gaia EDR3 parallax uncertainty in milliarcseconds.        ;
; gmag:     Gaia EDR3 G magnitude (original, not corrected).          ;
; ruwe:     Gaia EDR3 RUWE.                                           ;
; npar:     Gaia EDR3 number of parameters of the solution (5 or 6).  ;
;                                                                     ;
;;;;;;;;;;;;;;;;;;;;;;;;;;;;;;;;;;;;;;;;;;;;;;;;;;;;;;;;;;;;;;;;;;;;;;;
gref  = [ 6.50, 7.50, 8.50, 9.50,10.25,10.75,11.25,11.75,12.25,12.75, $
         13.25,13.75,14.25,14.75,15.25,15.75,16.25,16.75,17.25,17.75]
kref  = [ 2.62, 2.38, 2.06, 1.66, 1.22, 1.41, 1.76, 1.76, 1.90, 1.92, $
          1.61, 1.50, 1.39, 1.35, 1.24, 1.20, 1.19, 1.18, 1.18, 1.14]
k     = INTERPOL(kref ,gref ,6>gmag<18,/SPLINE)
geref = [ 6.00,12.50,13.50,14.50,15.50,16.50,17.50]
keref = [ 0.50, 0.50, 1.01, 1.28, 1.38, 1.44, 1.32]
k     = k*(1 + $
       (INTERPOL(keref,geref,6>gmag<18,/SPLINE) > 0.5)*(ruwe GT 1.4))
k     = k*(1 + 0.25*(npar EQ 6))
out  = SQRT((spi*k)^2 + 0.0103^2)
RETURN, out
END
\end{verbatim}
}
\label{spicor}
\end{table}

\begin{table*}
 \caption{Function that calculates the {\it Gaia}~EDR3 zero point for an array of values.}
{\scriptsize
\begin{verbatim}
FUNCTION ZPEDR3, gmag, nueff, lat, npar, ORIG=orig
;;;;;;;;;;;;;;;;;;;;;;;;;;;;;;;;;;;;;;;;;;;;;;;;;;;;;;;;;;;;;;;;;;;;;;;;;;;;;;;;
;                                                                              ;
; This function returns the parallax zero point correction for Gaia EDR3 in    ;
;  microarcseconds for an array of values. gmag, nueff, lat, and npar must     ;
;  be arrays with the same number of elements. By default, it applies the      ;
;  new zero point. The code uses VALID_NUM in the IDL astronomy library:       ;
;  https://idlastro.gsfc.nasa.gov/ftp/pro/misc/valid_num.pro                   ;
;                                                                              ;
; Positional parameters:                                                       ;
; gmag:     Gaia EDR3 G magnitude (original, not corrected).                   ;
; nueff:    Effective wavenumber, real for five-parameter solutions and        ;
;            pseudocolor for six-parameter solutions, in inverse microns.      ;
; lat:      Ecliptic latitude in degrees.                                      ;
; npar:     Number of parameters of the solution (2, 5, or 6).                 ;
;                                                                              ;
; Keyword parameters:                                                          ;
; ORIG:     Flag to use the Lindegren correction instead of the new one.       ;
;                                                                              ;
;;;;;;;;;;;;;;;;;;;;;;;;;;;;;;;;;;;;;;;;;;;;;;;;;;;;;;;;;;;;;;;;;;;;;;;;;;;;;;;;
IF NOT KEYWORD_SET(ORIG) THEN BEGIN ; New
 gcut5 = [   6.0 ,   7.4 ,   9.2 ,  10.8 ,  11.2 ,  11.8 ,  12.2 ,  12.9 ,  13.1 ,  15.9 ,  16.1 ,  17.5 ,  18.0 ,  19.0 ,  20.0 ,  21.0 ]
 q500  = [ -54.33,  -8.17, -27.11, -20.03, -33.38, -36.95, -16.99, -26.49, -37.51, -32.82, -33.18, -25.02, -22.88, -18.40, -12.65, -18.22] 
 q501  = [ -11.97, -10.06,  -7.60,  -2.78, -12.21, -11.55,  -3.67, -10.63,  +3.33,  +5.15,  -1.31,  +5.83,   0.40,  +5.98,  -4.57, -15.24] 
 q502  = [ +25.39, +24.12, +22.48, +10.48,  +5.51,  -1.65, +15.81, +21.77, +20.41,  +6.50,  +5.48,  +6.57,  -5.10,  -6.46,  -7.46, -18.54] 
 q510  = [ -31.1 , -13.4 ,  +9.3 , +11.6 ,-132.3 ,-158.7 ,-109.9 , -76.0 ,  -2.9 ,  -9.10, -56.8 , -39.2 , -46.5 ,   0   ,   0   ,   0   ] 
 q511  = [ +19.1 , +23.7 , +29.6 , +34.8 , -10.2 , +13.2 , +63.2 , +43.2 , +29.6 , +12.2 , -38.1 , -29.1 , -35.4 ,  +5.5 , +97.9 ,+128.2 ] 
 q520  = [ -2529., -2529., -2529., -2529., -2529., -2529., -3625., -4353., -1675., -1341., -1705., -1284.,  -896.,   0   ,   0   ,   0   ] 
 q530  = [   0   ,   0   ,   0   ,   0   ,   0   ,   0   ,   0   ,   0   , +32.1 ,+168.0 ,+112.1 ,+196.3 ,+126.5 ,   0   ,   0   ,   0   ] 
 q540  = [-358.1 ,-358.1 ,-358.1 , -78.6 ,+203.1 ,-155.3 ,-144.2 , +23.6 , +99.5 ,+129.3 ,+153.1 ,+218.0 ,+190.2 ,+276.6 ,   0   ,   0   ] 
ENDIF ELSE BEGIN                    ; Lindegren
 gcut5 = [   6.0 ,                  10.8 ,  11.2 ,  11.8 ,  12.2 ,  12.9 ,  13.1 ,  15.9 ,  16.1 ,  17.5 ,          19.0 ,  20.0 ,  21.0 ]
 q500  = [ -26.98,                 -27.23, -30.33, -33.54, -13.65, -19.53, -37.99, -38.33, -31.05, -29.18,         -18.40, -12.65, -18.22] 
 q501  = [  -9.62,                  -3.07,  -9.23, -10.08,  -0.07,  -1.64,  +2.63,  +5.61,  +2.83,  -0.09,          +5.98,  -4.57, -15.24] 
 q502  = [ +27.40,                 +23.04,  +9.08, +13.28,  +9.35, +15.86, +16.14, +15.42,  +8.59,  +2.41,          -6.46,  -7.46, -18.54] 
 q510  = [ -25.1 ,                 +35.3 , -88.4 ,-126.7 ,-111.4 , -66.8 ,  -5.7 ,   0   ,   0   ,   0   ,           0   ,   0   ,   0   ] 
 q511  = [  -0.0 ,                 +15.7 , -11.8 , +11.6 , +40.6 , +20.6 , +14.0 , +18.7 , +15.5 , +24.5 ,          +5.5 , +97.9 ,+128.2 ] 
 q520  = [ -1257.,                 -1257., -1257., -1257., -1257., -1257., -1257., -1189., -1404., -1165.,           0   ,   0   ,   0   ] 
 q530  = [   0   ,                   0   ,   0   ,   0   ,   0   ,   0   ,+107.9 ,+243.8 ,+105.5 ,+189.7 ,           0   ,   0   ,   0   ] 
 q540  = [   0   ,                   0   ,   0   ,   0   ,   0   ,   0   ,+104.3 ,+155.2 ,+170.7 ,+325.0 ,        +276.6 ,   0   ,   0   ] 
ENDELSE ; Six-parameter solutions use Lindegren
gcut6  = [   6.0 ,                  10.8 ,  11.2 ,  11.8 ,  12.2 ,  12.9 ,  13.1 ,  15.9 ,  16.1 ,  17.5 ,          19.0 ,  20.0 ,  21.0 ]
q600   = [ -27.85,                 -28.91, -26.72, -29.04, -12.39, -18.99, -38.29, -36.83, -28.37, -24.68,         -15.32, -13.73, -29.53] 
q601   = [  -7.78,                  -3.57,  -8.74,  -9.69,  -2.16,  -1.93,  +2.59,  +4.20,  +1.99,  -1.37,          +4.01, -10.92, -20.34] 
q602   = [ +27.47,                 +22.92,  +9.36, +13.63, +10.23, +15.90, +16.20, +15.76,  +9.28,  +3.52,          -6.03,  -8.30, -18.74] 
q610   = [ -32.1 ,                  +7.7 , -30.3 , -49.4 , -92.6 , -57.2 , -10.5 , +22.3 , +50.4 , +86.8 ,         +29.2 , -74.4 , -39.5 ] 
q611   = [ +14.4 ,                 +12.6 ,  +5.6 , +36.3 , +19.8 ,  -8.0 ,  +1.4 , +11.1 , +17.2 , +19.8 ,         +14.1 ,+196.4 ,+326.8 ] 
q612   = [  +9.5 ,                  +1.6 , +17.2 , +17.7 , +27.6 , +19.9 ,  +0.4 , +10.0 , +13.7 , +21.3 ,          +0.4 , -42.0 ,-262.3 ] 
q620   = [   -67.,                  -572., -1104., -1129.,  -365.,  -554.,  -960., -1367., -1351., -1380.,          -563.,  +536., +1598.] 
n      = N_ELEMENTS(gmag)
IF n NE N_ELEMENTS(nueff) OR n NE N_ELEMENTS(lat) OR n NE N_ELEMENTS(npar) THEN STOP, 'Incompatible data' 
c0     = 1.0 + FLTARR(n)
c1     =        -0.24*(nueff LE 1.24) +   (nueff-1.48)*(nueff GT 1.24 AND nueff LE 1.72) +         0.24*(nueff GT 1.72)
c2     =      +0.24^3*(nueff LE 1.24) + (1.48-nueff)^3*(nueff GT 1.24 AND nueff LE 1.48)
c3     = (nueff-1.24)*(nueff LE 1.24)
c4     =                                                                                   (nueff-1.72)*(nueff GT 1.72)
b0     = 1.0 + FLTARR(n)
b1     = SIN(!DTOR*lat)
b2     = SIN(!DTOR*lat)*SIN(!DTOR*lat) - 1./3
gmagc  = MIN(gcut5) > gmag < MAX(gcut5) ; Caps G between 6 and 21 
p      = WHERE(VALID_NUM(gmagc) EQ 0, np)
IF np  NE 0 THEN gmagc[p] = 21.0 ; Stars with non-valid G are assumed to be faint
qq500  = INTERPOL(q500,gcut5,gmagc)
qq501  = INTERPOL(q501,gcut5,gmagc)
qq502  = INTERPOL(q502,gcut5,gmagc)
qq510  = INTERPOL(q510,gcut5,gmagc)
qq511  = INTERPOL(q511,gcut5,gmagc)
qq520  = INTERPOL(q520,gcut5,gmagc)
qq530  = INTERPOL(q530,gcut5,gmagc)
qq540  = INTERPOL(q540,gcut5,gmagc)
qq600  = INTERPOL(q600,gcut6,gmagc)
qq601  = INTERPOL(q601,gcut6,gmagc)
qq602  = INTERPOL(q602,gcut6,gmagc)
qq610  = INTERPOL(q610,gcut6,gmagc)
qq611  = INTERPOL(q611,gcut6,gmagc)
qq612  = INTERPOL(q612,gcut6,gmagc)
qq620  = INTERPOL(q620,gcut6,gmagc)
z5     = qq500*c0*b0 + qq501*c0*b1 + qq502*c0*b2 + qq510*c1*b0 + qq511*c1*b1               + qq520*c2*b0 + qq530*c3*b0 + qq540*c4*b0
z6     = qq600*c0*b0 + qq601*c0*b1 + qq602*c0*b2 + qq610*c1*b0 + qq611*c1*b1 + qq612*c1*b2 + qq620*c2*b0
out    = z5*(npar EQ 5) + z6*(npar EQ 6) ; Returns zero for npar = 2 (or indeed any value other than 5 or 6)
RETURN, out
END
\end{verbatim}
}
\label{zpedr3}
\end{table*}

\end{document}